\documentclass[runningheads]{llncs}
\usepackage[table]{xcolor}
\usepackage{comment}
\DeclareMathAlphabet\mathbfcal{OMS}{cmsy}{b}{n}
\usepackage{amsmath}
\usepackage{amssymb}
\usepackage{algorithm}
\usepackage{algorithmicx}
\usepackage[noend]{algpseudocode}
\usepackage{xspace}
\usepackage{graphicx}
\usepackage{pifont}\usepackage{mathtools}
\usepackage{subcaption}
\usepackage[style=base,font=small]{caption}
\usepackage{mathrsfs}
\usepackage{booktabs} \usepackage{epstopdf}
\usepackage{comment}
\usepackage{tabularx}
\usepackage{wrapfig}
\usepackage{tabularx}
\usepackage{paralist}
\usepackage{balance}
\usepackage{todonotes}
\usepackage{bold-extra}  
\usepackage{multibib}
\newcites{latex}{Additional References}

\usepackage{hyperref}
\hypersetup{    colorlinks=true,
    linkcolor=mydarkblue,
    citecolor=mydarkblue,
    filecolor=mydarkblue,
    urlcolor=mydarkblue}
    
\usepackage{microtype}
\usepackage{balance}
   
\usepackage{amsmath,amssymb,amsthm}

\usepackage{tikz}
\usepackage{verbatim}
\usetikzlibrary{arrows}
\usetikzlibrary{shapes}
\usetikzlibrary{decorations.pathmorphing} \usetikzlibrary{fit}					\usetikzlibrary{backgrounds}	
\usepackage{ragged2e}
\usepackage{multirow}
\usepackage{microtype}
\usepackage{balance}
\usepackage{setspace}
\usepackage{etoolbox}

\makeatletter
\newcommand*\bigcdot{\mathpalette\bigcdot@{.5}}
\newcommand*\bigcdot@[2]{\mathbin{\vcenter{\hbox{\scalebox{#2}{$\m@th#1\bullet$}}}}}
\makeatother

\makeatletter
\newcommand*{\algrule}[1][\algorithmicindent]{  \makebox[#1][l]{    \hspace*{.2em}    \vrule height .75\baselineskip depth .25\baselineskip
  }
}
\newcount\ALG@printindent@tempcnta
\def\ALG@printindent{    \ifnum \theALG@nested>0    \ifx\ALG@text\ALG@x@notext        \else
    \unskip
        \ALG@printindent@tempcnta=1
    \loop
    \algrule[\csname ALG@ind@\the\ALG@printindent@tempcnta\endcsname]    \advance \ALG@printindent@tempcnta 1
    \ifnum \ALG@printindent@tempcnta<\numexpr\theALG@nested+1\relax
    \repeat
    \fi
    \fi
}
\patchcmd{\ALG@doentity}{\noindent\hskip\ALG@tlm}{\ALG@printindent}{}{\errmessage{failed to patch}}
\patchcmd{\ALG@doentity}{\item[]\nointerlineskip}{}{}{} \makeatother

\algnewcommand\algorithmicforeach{\textbf{for each}}
\algdef{S}[FOR]{ForEach}[1]{\algorithmicforeach\ #1\ \algorithmicdo}

\newcolumntype{C}[1]{>{\centering\let\newline\\\arraybackslash\hspace{0pt}}m{#1}}
\newcolumntype{R}[1]{>{\RaggedLeft\arraybackslash}} \newcolumntype{L}[1]{>{\RaggedRight\arraybackslash}} 

\newcommand{\eat}[1]{\ignorespaces}

\graphicspath{{./}{./graphics/}}
\newcolumntype{H}{>{\setbox0=\hbox\bgroup}c<{\egroup}@{}}

\newcommand{\eg}{\emph{e.g.}}
\newcommand{\ie}{\emph{i.e.}}

\newtheorem{Definition}{\bfseries{Definition}}
\newtheorem{Problem}{\bfseries{Problem}}

\providecommand{\mat}[1]{\boldsymbol{\mathrm{#1}}}\renewcommand{\vec}[1]{\boldsymbol{\mathrm{#1}}}

\DeclareMathOperator{\hugeE}{\mbox{\huge\raise-0.3ex\hbox{E}}}
\DeclareMathOperator{\p}{\mathbb{P}}
\DeclareMathOperator{\hugep}{\mbox{\huge\raise-0.3ex\hbox{$\p$}}}

\newcommand{\RR}{\mathbb{R}}

\providecommand{\nodetype}{\ensuremath{p}}

\providecommand{\mF}{\ensuremath{\mat{F}}}

\providecommand{\mZ}{\ensuremath{\mat{Z}}}

\providecommand{\vh}{\ensuremath{\vec{h}}}

\providecommand{\vr}{\ensuremath{\vec{r}}}

\providecommand{\vz}{\ensuremath{\vec{z}}}

\newtheorem{Result}{\textbf{\textsc{Conclusion}}}

\usepackage{numprint}

\newcommand{\rr}[1]{{\textcolor{purple}{#1}}}

\newcommand{\dijin}[1]{{\textcolor{blue}{#1}}}

\newcommand{\method}{\textsc{node2bits}\xspace} \newcommand{\abbre}{\textsc{n2b}\xspace}
\newcommand{\abbrebase}{\textsc{n2b-0}\xspace}
\newcommand{\abbreshort}{\textsc{n2b-sh}\xspace}
\newcommand{\abbrelong}{\textsc{n2b-ln}\xspace}
\newcommand{\abbreunsup}{\textsc{n2b-u}\xspace}
\newcommand{\methodbase}{\textsc{node2bits-0}\xspace}
\newcommand{\methodshort}{\textsc{node2bits-sh}\xspace}
\newcommand{\methodlong}{\textsc{node2bits-ln}\xspace}
\newcommand{\methodunsup}{\textsc{node2bits-u}\xspace}

\newcommand\TT{\rule{0pt}{2.7ex}}

\newcommand{\deltat}{{\Delta t}}

\usepackage{url}
\urldef{\mailsa}\path|{dijin, mheimann, dkoutra}@umich.edu|

\titlerunning{Compact Time- and Attribute-aware Node Representation Learning}

\begin{document}

\title{\method: Compact Time- and Attribute-aware Node Representations for User Stitching}

\author{Di Jin\inst{1} \and Mark Heimann\inst{1} \and Ryan A. Rossi\inst{2}\and Danai Koutra\inst{1}
}
\institute{University of Michigan, Ann Arbor 
\and Adobe Research
}
\maketitle

\begin{abstract}
Identity stitching, the task of identifying and matching various online references (\eg, sessions over different devices and timespans) to the same user in real-world web services, is crucial for personalization and recommendations. 
However, traditional user stitching approaches, such as grouping or blocking, require pairwise comparisons between a massive number of user activities, thus posing both computational and storage challenges. Recent works, which are often application-specific, heuristically seek to reduce the amount of comparisons, but they suffer from low precision and recall.  To solve the problem in an application-independent way, we take a heterogeneous network-based approach in which users (nodes) interact with content (\eg, sessions, websites), and may have attributes (\eg, location). We propose \method, an efficient framework that represents multi-dimensional features of node contexts with binary hashcodes.
\method leverages \textit{feature-based temporal} walks to encapsulate short- and long-term interactions between nodes in heterogeneous web networks, and adopts SimHash to obtain compact, binary representations and avoid the quadratic complexity for similarity search.
Extensive experiments on large-scale real networks show that \method outperforms traditional techniques and existing works that generate real-valued embeddings by up to $5.16\%$ in $F1$ score on user stitching, while taking only up to $1.56\%$ as much storage. 
\end{abstract}

\definecolor{mydarkblue}{rgb}{0,0.08,0.45} 
\vspace{-0.6cm}
\section{Introduction}
\vspace{-0.1cm}
\label{sec:intro}
Personalization and  recommendations increase user satisfaction by providing relevant experiences and handling the online information overload in news, web search, entertainment, and more. Accurately modeling user behavior and preferences over time are at the core of personalization. However, tracking user activity online is challenging as users interact with tens of internet-enabled devices from different locations daily, leading to fragmented user profiles. 
Without unified  profiles, the observed user data are sparse, non-representative of the population, and insufficient for accurate predictions that drive business success.

In this work, we tackle the problem of \textit{identity or user stitching}, which aims to identify and group together logged-in and anonymous sessions that correspond to the \textit{same} user despite taking place across different channels, platforms, devices and browsers~\cite{saha2015probabilistic}. 
This problem is a form of entity or identity resolution~\cite{getoor2013entity,bhattacharya:07}, also known as entity linking, record linkage, and duplicate detection~\cite{Christen12,Kolb12_dedoop,bhattacharya:07}. Unlike entity resolution where textual information per user (\eg, name, address) is available, identity stitching 
relies solely on user interactions with online content and web metadata. Although cookies can help stitch several different sessions of the same user, many users have multiple cookies ({\eg, a cookie for each device or web browser})~\cite{dasgupta2012overcoming}, and most cookies expire after a short time, and therefore cannot help to stitch users over time.
Similarly, IP addresses change across locations resulting in fragmentation or even erroneous stitching between users who have the same IP address at different times (e.g., airports).  Meanwhile, fingerprinting approaches~\cite{eckersley2010unique} capture user similarity based on device or browser configurations, not on behavioral patterns that remain consistent across devices or browsers.  
On the other hand, exhaustive solutions for entity resolution require quadratic number of comparisons between all pairs of entities, which is computationally intractable for large-scale web services. 
This can be partially handled via the heuristic of blocking~\cite{Papadakis16_blocking}, which groups similar entity descriptions into blocks, and only compares entities within the same block.

To overcome these challenges and better tailor to the user stitching setup, our solution is based on the idea that the same user accesses \textit{similar content} across platforms and has \textit{similar behavior} over time. We model the user interactions with different content and platforms over time in a dynamic heterogeneous network, where user stitching maps to the identification of \textit{nodes} that correspond to the same real-world entity.
Motivated by the success of node representation learning, we aim to find embeddings of time-evolving `user profiles' over this rich network of interactions.  
For large-scale graphs, however, the customary dense node representations for each node can often impose a formidable memory requirement, on par with that of the original (sparse) adjacency matrices~\cite{JinRKKRK19multilens}.  
Thus, to efficiently find \emph{sparse} binary representations and link entities based on similar activity while avoiding the pairwise comparison of all user profiles, 
we solve the following problem:

\begin{Problem}[Temporal, Hash-based Node Embeddings] \label{problem}
	Given a graph $G(V,E)$, the goal of \textit{hash-based network embedding} is to learn a function $\chi: V \rightarrow \{0,1\}^d$ such that the derived \emph{binary} $d$-dimensional embeddings  (1)~preserve similarities in interactions in $G$, 
	(2)~are space-efficient, and 
	(3)~accurately capture temporal information and the heterogeneity of the underlying network.
\end{Problem}

\begin{wrapfigure}{r}{0.66\textwidth}
	\vspace{-0.9cm}
	\centering
	\includegraphics[width=\linewidth]{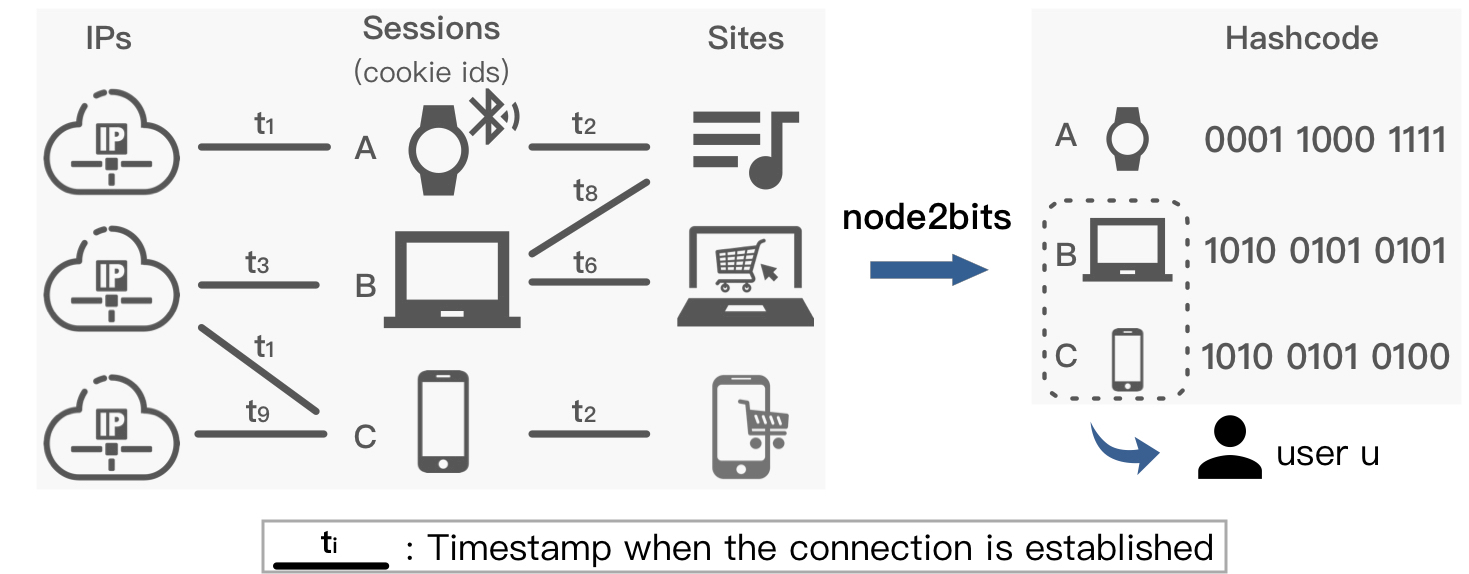}
	\vspace{-0.4cm}
	\caption{{\method overview. \method encodes the temporal, heterogeneous information of each node into binary hashcodes for efficient user stitching.}
	}
	\label{fig_illustration}
	\vspace{-0.8cm}
\end{wrapfigure}
We introduce a \textit{general} framework called \method that captures temporally-valid interactions between nodes in a network, and constructs the contexts based on topological features and (optional) side information of entities involved in the interaction.
These feature-based contexts are then turned into histograms that incorporate node type information at different \textit{temporal distances}, and are mapped to binary hashcodes through SimHash~\cite{charikar2002similarity}. 
Thanks to locality sensitive hashing~\cite{indyk1998approximate}, the hashcodes, which are time-, attribute- and structure-aware, preserve the similarities in temporal interaction patterns in the network, and achieve both space and computational efficiency for similarity search. 
Given these sparse, hash-based embeddings of all entities, we then cast user stitching as a supervised binary classification task or a hashing-based unsupervised task.
As an example, in Fig.~\ref{fig_illustration}, devices $B$ and $C$ are associated with identical IPs and similar online sales websites visited afterwards, thus they are encoded similarly and could correspond to the same user.

Our contributions are:

\vspace{-0.2cm}
\begin{itemize}
	\item {\bf Embedding-based Formulation:} Going beyond traditional blocking techniques, we formulate the problem of user stitching as the problem of finding {temporal, hash-based embeddings in heterogeneous networks} such that they maintain \textit{similarities between user interactions over time}.
	
	\item {\bf Space-efficient Embeddings:} We propose \method, a practical, intuitive, and fast framework that generates \textit{compact, binary embeddings} suitable for user stitching. Our method combines random walk-based sampling of contexts, their feature-based histogram representations, and locality sensitive hashing to preserve the heterogeneous equivalency of contexts over time.         
	
	\item {\bf Extensive Empirical Analysis:} Our experiments on real-world networks show that \method outputs a space-efficient binary representation which uses between $63\times$ and $339\times$ less space than the baselines
	while achieving comparable or better performance in user stitching tasks. 
	Moreover, \method is scalable for large real-world temporal and heterogeneous networks.

\end{itemize}
For reproducibility, the code is at \url{https://github.com/GemsLab/node2bits}.

\vspace{-0.2cm}
\section{Preliminaries and Definitions}
\label{sec:prelim}
Before we introduce \method, we discuss two key concepts that our method is based on: our dynamic heterogeneous network model, and temporal random walks.  We give the main symbols and their definitions in Table~\ref{table_symbols}. 
\begin{table}[t]
	\centering
	\vspace{-0.3cm}
	\caption{Summary of major symbols and their definitions. }
	\centering 
	\fontsize{7}{7.5}\selectfont
	\setlength{\tabcolsep}{5pt} \label{table_symbols}
	\def\arraystretch{1.35} \begin{tabularx}{1.0\linewidth}{@{}rX@{}}
		\toprule
		\textbf{Symbol} & \textbf{Definition} 
		\\ 
		\midrule
		
		$G(V, E, \xi, \psi)$ & (un)directed and (un)weighted heterogeneous network with nodeset $V$, edgeset $E$, a mapping $\xi$ from nodes to node types, and an edge mapping $\psi$, resp. \\
		$|V|=N, |E|=M$ & number of nodes and edges in $G$ \\
		$\mathcal{T}_{V}, |\mathcal{T}_{V}| $\;$ ; $\;$ \mathcal{T}_{E}, |\mathcal{T}_{E}|$ & set of node/edge types in the heterogeneous graph and its size, resp. \\
		
		$\mF$ & $N \times |\mathcal{F}|$ feature matrix including node attributes and derived features \\
		$f_{ij}$, $f_{(j)}$ & $(i,j)^{th}$ element of $\mF$ and index of its $j^{th}$ feature, resp.\\
		$\mathcal{W}$ & set of random walks \\
		$(\mathbf{w}_L)_{L\in\mathbb{N}}, \mathbf{w}_L[u]$ & sequence of nodes in a random walk of length $L$, and the index of node $u$, resp. \\
		$L$ & the maximum length of a random walk \\
		
		$\deltat$ & `temporal distance' in $\mathcal{W}$ based on temporally ordered edge transitions\\
		$\mathcal{C}^{\deltat}_u$, $\mathcal{C}^{\deltat}_u|f$ & context of node $u$ at distance $\deltat$, and feature-based context, resp. \\
		$g_i: \mathcal{C} \rightarrow \{0,1\}$ & $i^{th}$ LSH function that hashes a node context into a binary value \\
		$K^{\deltat}$, $K$ & embedding dimension at distance $\deltat$, and output dimension $K = \sum_{\deltat=1}^{\text{MAX}}K^{\deltat}$\\
		$\vh(\mathcal{S}), \vh(\mathcal{S}| \cdot)$ & unconditional and conditional $b$-bin histogram of values in enclosed set $\mathcal{S}$, resp. \\ 
		$\mZ$ & $N \times K$ output binary embeddings or hashcodes \\
		\bottomrule
	\end{tabularx}
	\vspace{-0.4cm}
\end{table}

\subsection{Dynamic Heterogeneous Network Model} \label{sec:heter-graph-model}

As we mentioned above, we model the user interactions with content, websites, devices etc. as a heterogeneous network, which is formally defined as: 
\begin{Definition}[\sc Heterogeneous Nework] \label{def:hetero-graph}
	\vspace{-0.2cm}
	A heterogeneous network  $G=(V,E,\psi,\xi)$ is comprised of 
	(i) a nodeset $V$ and edgeset $E$, (ii) a mapping $\psi:V\rightarrow\mathcal{T}_V$ of nodes to their types, and (iii) a mapping $\xi:E\rightarrow\mathcal{T}_E$ to edge types. \end{Definition}\noindent

Many  graph types are  special cases of heterogeneous networks:  
({\bf 1}) homogeneous graphs have $|\mathcal{T}_V|=|\mathcal{T}_E|=1$ type;  
({\bf 2}) $k$-partite graphs consist of $|\mathcal{T}_V|=k$ and $|\mathcal{T}_E|=k-1$ types; 
({\bf 3}) signed networks have $|\mathcal{T}_V|=1$ and $|\mathcal{T}_E|=2$ types; and
({\bf 4}) labeled graphs have a single label per node/edge. 
Most real networks capture evolving processes (e.g., communication, browsing activity) and thus change over time.
Instead of approximating a dynamic network as a sequence of lossy discrete static snapshots $G_1, \ldots, G_T$, 
we model the \emph{temporal interactions} in a \textit{lossless} fashion as a \emph{continuous-time dynamic network}~\cite{nguyen2018continuous}.
\begin{Definition}[\sc Continuous-Time Dynamic Network] \label{eq:cont-time-dynamic-network}
	\vspace{-0.1cm}
	A continuous-time dynamic, heterogeneous network $G=(V,E_\tau,\psi, \xi, \tau)$ is a heterogeneous network with $E_\tau$ temporal edges between vertices $V$, 
	where $\tau: E \rightarrow \RR^{+}$ is a function that maps each edge to a corresponding timestamp.
\end{Definition}\noindent

\vspace{-0.35cm}
\subsection{Temporal Random Walks } \label{sec:temporally-valid}
A walk on a graph is a sequence of nodes where each pair of successive nodes are connected by an edge.  
Popular network embedding methods generate walks using randomized procedures \cite{deepwalk,node2vec} to construct a corpus of node IDs or node context. 
In continuous-time dynamic networks, a \emph{temporally valid} walk is defined as a sequence of nodes connected by edges with non-decreasing timestamps (\eg, representing the order that user-content interactions occurred) and were first proposed and used for embeddings in~\cite{nguyen2018continuous}.
\begin{Definition}[\sc Temporal Walk]\label{def:temporal-walk}
	A temporal walk of length $L$ from $v_1$ to $v_L$ in graph $G=(V,E,\psi,\xi)$ is a sequence of vertices $\langle v_1, v_2, \cdots, v_L \rangle$ such that $\langle v_i, v_{i+1} \rangle \in E_\tau$ for $1 \leq i < L$, and the timestamps are in valid temporal order: $\tau(v_i, v_{i+1}) \leq \tau(v_{i+1}, v_{i+2})$ for $1 \leq i < (L-1)$. 
\end{Definition}
\noindent

\vspace{-0.35cm}
\section{\method: Hash-based Emdedding Framework} 
\label{sec:approach}

\begin{figure*}[!t]
	\centering
	\includegraphics[width=\linewidth]{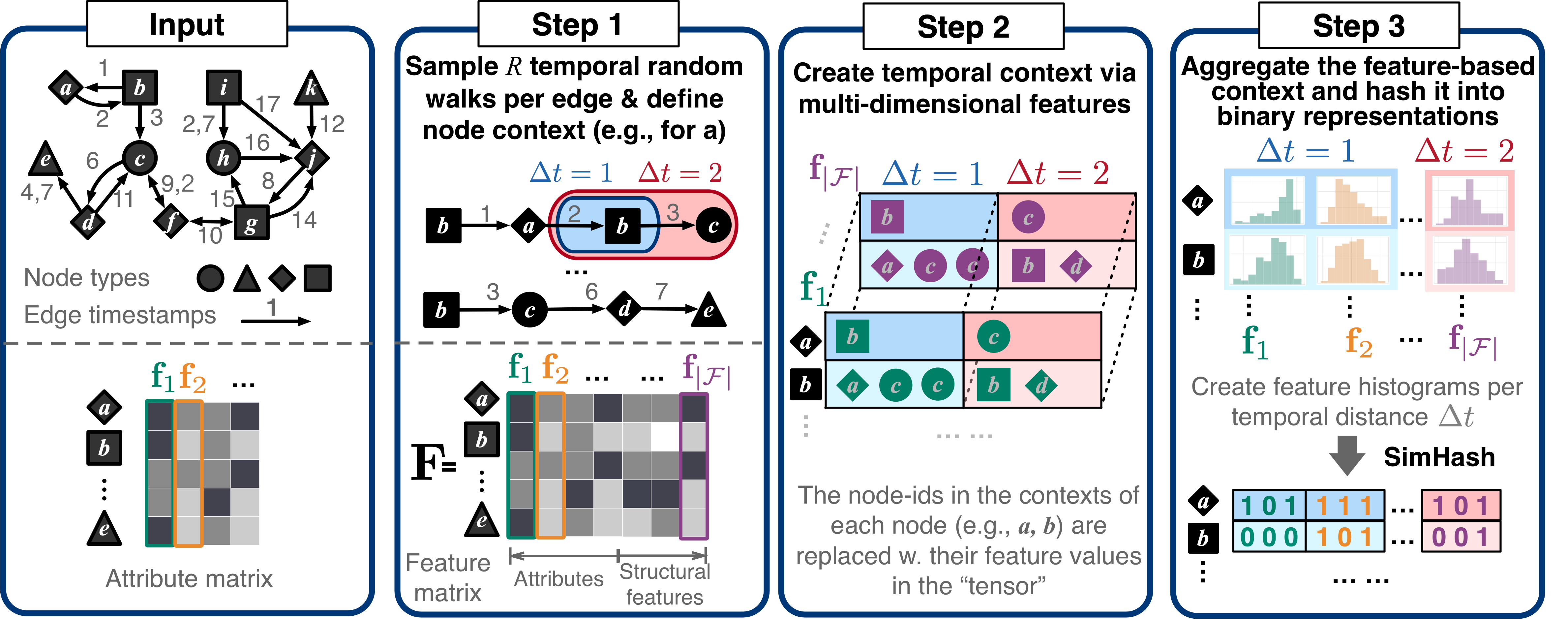}
	\caption{\method workflow. Given a graph and its attribute matrix  (optional), \method 
		(1)~samples \textit{temporal} random walks to obtain sequences that respect time,
		derives contexts at different temporal distances (temporal contexts of {\em a} and {\em b} are derived from the walk $\{b, a, b, c\}$, 
		as well as the feature matrix $\mathbf{F}$;
		(2)~creates temporal contexts based on multi-dimensional features in $\mF$; and 
		(3)~aggregates them into feature-based histograms to obtain sparse, binary, similarity-preserving embeddings via SimHash. }
	\label{fig_workflow}
	\vspace{-0.6 cm}
\end{figure*}

Motivated by the task of user stitching, we aim to develop \method to compactly describe each node/entity in the context of \textit{realistic interactions} (Problem~\ref{problem}).
Accordingly, \method is required to: 
\textbf{(R1)} support heterogeneous networks where the nodes and edges can be of any arbitrary type (\eg, a user, web page, IP, tag, spatial location); 
\textbf{(R2)} preserve the temporal validity of the events and interactions in the data; 
\textbf{(R3)} scale in runtime to large networks with millions of nodes/edges; and \textbf{(R4)} scale in memory requirements with space-efficient yet powerful \eat{sparse} \emph{binary} embeddings that capture ID-independent similarities. 
Next we detail the three main steps of \method: 
(\S~\ref{sec:RW}) Sampling temporal random walks and defining temporal contexts; 
(\S~\ref{sec:feat_rw}) Constructing temporal contexts based on multi-dimensional features;  (\S~\ref{sec:aggr-hashing}) Aggregating and hashing contexts into sparse embeddings.
We give the overview of \method in Figure~\ref{fig_workflow} and Algorithm~\ref{alg_workflow}.

\vspace{-0.2cm}
\subsection{Temporal Random Walk Sampling}
\label{sec:RW}
The first step of \method is to capture interactions in a node's context, which is important for the user stitching task: instead of simple interactions corresponding to pairwise edges, it samples more complex interaction sequences via random walks. But unlike many existing representation learning approaches~\cite{deepwalk,node2vec},  our method samples \textit{realistic} interactions in the order that they happen via $L$-step \textit{temporal} random walks (Definition \ref{def:temporal-walk}~\cite{nguyen2018continuous}), thus satisfying requirement \textbf{R2}. 
\method defines the \textit{temporal context} $\mathcal{C}_u^{\deltat}$ of node $u$ at temporal distance $\deltat$ as the \textit{collection of entities} that are at $\deltat$-hops away from node $u$ in the sampled random walks. Formally: \begin{equation}
\mathcal{C}^{\deltat}_u = \{v \; : \;  |\mathbf{w}_L[v]-\mathbf{w}_L[u]|=\deltat, \; \forall{\mathbf{w}_L}\in\mathcal{W}\},
\label{eq_context}
\end{equation}
where $\mathbf{w}_L[\cdot]$ is the index of the corresponding node in the random walk $(\mathbf{w}_L)_{L\in\mathbb{N}}$. For example, in Figure~\ref{fig_workflow} (Step 1) the {context of node $a$ at temporal distance $2$ is $\mathcal{C}^{\deltat=2}_a = \{c\}$ (highlighted in red). }Depending on the temporal context that we want to capture, $\deltat$ can vary up to a $MAX$ distance. 
Intuitively, small values of temporal distance capture more \textit{direct} interactions and similarities between entities. 
In static graphs, $\deltat$ simply corresponds to the distance between nodes in the sampled sequences, without capturing any temporal information.

\medskip\noindent\textbf{Temporal locality.}
The context that is defined above does not explicitly incorporate the \textit{time elapsed} between consecutively sampled interactions.
However, when modeling temporal user interactions, it is important to distinguish between short-term and long-term transitions.
Inspired by~\cite{nguyen2018continuous}, \method accounts for the \textit{closeness} or \textit{locality} between consecutive contexts (\ie, $\mathcal{C}^{\deltat}_u$ and $\mathcal{C}^{\deltat + 1}_u$)
through different biased temporal walk policies. For example, in the short-term policy, the transition probability from node $u$ to  $v$ is given as the softmax function: \begin{equation}
\small
P[v|u] = \frac{\exp{({-\tau(u,v)}/{d})}}{ \sum_{i\in\Gamma_\tau(u)}\exp{({-\tau(u,i)}/{d})} }
\label{eq_trans_short}
\end{equation}
where $\tau()$ maps an edge to its timestamp, $d=\max_{e\in E_\tau}{\tau(e)} - \min_{e\in E_\tau}{\tau(e)}$ is the total duration of all timestamps, and $\Gamma_\tau(u)$ is the set of temporal neighbors reached from node $u$ through temporally valid edges. Similarly, in the long-term policy, the transition probability from node $u$ to $v$ is given as in Equation~\eqref{eq_trans_short} but with positive signs in the numerator and denominator.

\vspace{-0.2cm}
\subsection{Temporal Context based on Multi-dimensional Features} \label{sec_multi_context}
\label{sec:feat_rw}

The context in Equation~\eqref{eq_context} depends on the node identities (IDs). 
However, in a multi-platform environment, a single entity may have multiple node IDs, thus contributing to seemingly different contexts. To generate ID-independent contexts that are appropriate for user stitching, we make the temporal contexts attribute- or feature-aware (\textbf{R1}), by building on the assumption that corresponding or similar entities have similar features.
Formally, we assume that a network may have a set of input node attributes (\eg, IP address, device type), as well as a set of derived topological features (\eg, degree, PageRank), all of which are stored in a $N \times |\mathcal{F}|$ feature matrix $\mF$ (Figure~\ref{fig_workflow}, Step 1).
We then generalize our random walks to not only respect time (\textbf{R2})~\cite{nguyen2018continuous}, but also capture this feature information using the notion of attributed/feature-based walks proposed in~\cite{role2vec}: 

\begin{Definition}[\sc Feature-based Temporal Walk] 
	\label{def:attr-temporal-random-walk}
	A \emph{feature-based temporal walk} of length $L$ from node $v_1$ to $v_L$ in graph $G$ is defined as a sequence of feature values corresponding to the sequence of vertices in a valid temporal walk (Dfn.~\ref{def:temporal-walk}). For the $j^{th}$ feature $f_{(j)}$, the corresponding feature-based temporal walk is 
	\vspace{-0.1cm}
	\begin{gather} \label{eq:attr-random-walk}
		\langle w_{L,f_{(j)}} \rangle_{L\in\mathbb{N}} = \langle f_{v_1,j}, f_{v_2,j},\ldots, f_{v_L,j} \rangle,
	\end{gather}
	\vspace{-0.1cm}
	where $f_{v_i,j}$ is the value of the $j^{th}$ feature for node $v_i$, stored in matrix $\mF$. 
\end{Definition}\noindent
Our definition is general as it allows walks to obey time while  each node may have a $d$-dimensional vector of input attribute values and/or derived structural features, which can be discrete or real-valued~\cite{role2vec}.

\vspace{-0.3cm}
\subsubsection{Temporally-valid, multi-dimensional feature contexts.} 
\method extends the previously generated temporal contexts to incorporate node features and remove the dependency on node IDs. Following the definition of feature-based temporal walks, given $|\mathcal{F}|$ features, our method generates $|\mathcal{F}|$-dimensional contexts per node $u$ and temporal distance $\deltat$ by replacing the node IDs in Equation~\eqref{eq_context} with their corresponding feature values (Figure~\ref{fig_workflow}, Step 2). Formally, the temporally-valid, multi-dimensional feature contexts are defined as: 
\begin{equation}
\mathcal{C}^{\deltat}_u | f_{(j)} = \{f_{v, j} \; : \; \forall{v}\in\mathcal{C}^{\deltat}_{u}\} \; \; \forall \text{ feature } f_{(j)} \in \mathcal{F},
\label{eq_context_feature}
\end{equation}
where $f_{v,j}$ is the value of the $j^{th}$ feature for node $v$.

\vspace{-0.2cm}
\subsection{Feature-based Context Aggregation and Hashing} \label{sec:aggr-hashing}

The key insight in user stitching is that each user interacts with similarly `typed' entities through similar relations over time: for example, in online-sales logs, a user likely browses similar types of goods in logged-in and anonymous sessions; and in online social networks, accounts sharing near-identical interaction patterns, such as replies or shares, are potentially from the same person. Based on this insight, \method augments the previously generated temporal, multi-dimensional feature contexts with node types (and implicitly the corresponding relations or edge types), which is a key property of heterogeneous networks (\textbf{R1}). It subsequently aggregates them and derives similarity-preserving and space-efficient, binary entity representations (\textbf{R4}) via locality sensitive hashing. 

\vspace{-0.3cm}
\subsubsection{Context Aggregation.} Unlike existing works that aggregate contextual features into a single value such as mean or maximum~\cite{hamilton2017inductive,rossi2018deep},  \method aggregates them into \textit{less lossy} representations: 
\textit{histograms} tailored to heterogeneous networks by distinguishing between node types (\textbf{R1}). Specifically, it models the transitional dependency across node and relation types by further conditioning the derived contexts in Equation~\eqref{eq_context_feature} on the node types $p_i\in \mathcal{T}_V$ (\ie, each temporal context consists of the features of only one node type). 
We denote the temporal contexts conditioned on both features and node types as $\mathcal{C}_{u}^{\deltat}|f, p$. The final histogram representation of node $u$ at temporal distance $\deltat$ consists of the concatenation of the histograms over the conditional contexts at $\deltat$ (Figure~\ref{fig_workflow}, Step 3): \begin{equation}
\vh(\mathcal{C}^{\deltat}_u)=[\textcolor{gray}{\vh(\mathcal{C}^{\deltat}_u\;|\;f_{(1)}, p_1)}, \textcolor{gray}{\vh(\mathcal{C}^{\deltat}_u\;|\;f_{(2)}, p_1)},\ldots,\textcolor{gray}{\vh(\mathcal{C}^{\deltat}_u\;|\;f_{(|\mathcal{F}|)}, p_{|\mathcal{T}_V|})}].
\label{eq_aggregation}
\end{equation}
In this representation, the features are binned logarithmically to account for the often skewed distributions of structural features (\eg degree).
We note that the histograms can be further extended to edge types as shown in~\cite{JinRKKRK19multilens}, for example by distinguishing pairs of nodes that are connected by multiple types of edges.

\vspace{-0.3cm}
\subsubsection{Similarity-preserving Representations via Hashing.} \label{sec_lsh}
Locality sensitive hashing (LSH) has been widely used for searching nearest neighbors in large-scale data mining~\cite{rajaraman2014mining}. In this work, we adopt SimHash~\cite{charikar2002similarity} to obtain similarity-preserving and space-efficient representations (\textbf{R4}) for all the entities in the heterogeneous network based on their aggregated time-, attribute-, and node type-aware contexts given by Equation~\eqref{eq_aggregation}.

Given a node-specific histogram $\vh(\mathcal{C}^{\deltat}_u) \in \mathbb{R}^{d}$ (with dimensionality $d= |\mathcal{F}||\mathcal{T}_V| \cdot b$, and $b$ being the number of logarithmic bins for the features),
SimHash generates a $K^{\deltat}$-dimensional\footnote{We assume that the length of each sketch at distance $\deltat$ is given as  $K^{\deltat}=\frac{K}{\text{MAX}}$.}
binary hashcode or sketch $\vz_u^{\deltat}$ by projecting the histogram to $K^{\deltat}$  
random hyperplanes  $\vr_i \in \mathbb{R}^d$ as follows: \begin{align}
	g_i(\; \vh(\mathcal{C}^{\deltat}_u) \;) = \text{sign} \left( \vh(\mathcal{C}^{\deltat}_u) \cdot \vr_i \right) 
	\label{eq_hashing}
\end{align}
In practice, the hyperplanes do not need to be chosen uniformly at random from a multivariate normal distribution, but it suffices to choose them uniformly from $\{-1,1\}^d$. The important property of locality sensitive hashing that guarantees that the similarities in the histogram space (which captures the temporal interactions between entities in $G$) are maintained is the following: for the SimHash family, the probability that a hash function agrees for two different vectors is equal to their cosine similarity. More formally, for two nodes $u$ and $v$:
\begin{align}
	P(\; g_i(\vh(\mathcal{C}^{\deltat}_u)) = g_i(\vh(\mathcal{C}^{\deltat}_v)) \;)  = 1 -   \frac{\cos^{-1} \frac{\vh(\mathcal{C}^{\deltat}_u) \cdot \vh(\mathcal{C}^{\deltat}_v)}{ |\vh(\mathcal{C}^{\deltat}_u)| |\vh(\mathcal{C}^{\deltat}_v)|}}{180}.
	\label{eq_cosine}
\end{align}
In other words, the cosine similarity between nodes $u$ and $v$ in the context-space is projected to the sketch-space and can be measured by the cardinality of matching between $\vz_u^\deltat$ and $\vz_v^\deltat$, where  
$\vz_{\bullet}^{\deltat} = [g_1(\; \vh(\mathcal{C}^{\deltat}_{\bullet}) \;), g_2(\; \vh(\mathcal{C}^{\deltat}_{\bullet}),  \ldots, g_{K^{\deltat}}(\; \vh(\mathcal{C}^{\deltat}_{\bullet})].$

For each node $u$ in $G$, the final binary representation is obtained by concatenating the hashcodes for contexts at different temporal distances $\deltat$, resulting in a $K$-dimensional vector (since $K = \sum_{\deltat=1}^{\text{MAX}}K^{\deltat}$):
\begin{align}
	\vz_u &= [\vz_u^{\deltat=1} \hspace{0.1cm} \vz_u^{\deltat=2} \hspace{0.1cm} \ldots \hspace{0.1cm} \vz_u^{\deltat=\text{MAX}} ]             \label{eq_z_representation}
\end{align}
where we replace the $-1$ bits with $0$s to achieve a more space-efficient representation (\textbf{R4}). An example is shown in the second half of Step 3 in Figure~\ref{fig_workflow}, where the blue shades denote histograms and sketches for contexts in temporal distance $\deltat=1$, and red shades correspond to $\deltat=2$.
Thus, the $K$-dimensional representation for each node, $\vz_u \in \{0,1\}^K$, captures the similarities between time-, feature- and node type-aware histograms across multiple temporal distances $\deltat$.  
The similarity between two nodes' histograms can be quickly estimated as the proportion of common bits in their binary representations $\vz_{\bullet}$.

\definecolor{lightgray}{rgb}{0,0.08,0.45} 
\algblockdefx[parallel]{ParFor}{EndPar}[1][]{$\textbf{parallel for}$ #1 $\textbf{do}$}{$\textbf{end parallel}$}
\algrenewcommand{\alglinenumber}[1]{\fontsize{6.5}{7}\selectfont#1}
\algtext*{EndPar}

\algblockdefx[parallel]{parfor}{endpar}[1][]{$\textbf{parallel for}$ #1 $\textbf{do}$}{$\textbf{end parallel}$}
\algrenewcommand{\alglinenumber}[1]{\scriptsize#1:}

\algblockdefx[foreach1]{foreach}{endforeach}[1][]{$\textbf{for}$ #1 $\textbf{do}$}{$\textbf{end for}$}
\algrenewcommand{\alglinenumber}[1]{\scriptsize#1}

\newcommand{\multilinenospaceD}[1]{\State \parbox[t]{\dimexpr0.94 \linewidth-\algorithmicindent}{\begin{spacing}{0.8}\fontsize{8}{9}\selectfont#1\strut \end{spacing}}}

\begin{algorithm}[t!]
	\scriptsize
	\caption{\method Framework 
	}
	\label{alg_workflow}
	{\begin{spacing}{1.15}
			\begin{algorithmic}[1]
				\fontsize{8}{9}\selectfont
				\Require (un)directed heterogeneous graph $G(V, E, \psi,\xi)$,
				\# random walks $R$ per edge, 
				max walk length $L$,
				max temporal distance $\text{MAX}$,
				embedding dimensionality $K^\deltat$ at dist.\ $\deltat$ 
				\medskip
				
				\State For each edge $e$, perform $R$ temporal walks based on the short- or long-term policy (\S~\ref{sec:RW})
				\State Obtain temporal contexts $\mathcal{C}^{\deltat}_u$ for each node $u$ at temporal distances $\deltat \leq \text{MAX}$ via Eq.~\eqref{eq_context} \State Construct feature matrix $\mF$ with node attributes (if avail.) and topological features (\S~\ref{sec:feat_rw}) \State Derive feature-based temporal contexts $\mathcal{C}^{\deltat}_u|f_{(j)}$ by replacing $v \in \mathcal{C}^{\deltat}_u$ with the feature value $f_{v,j}$, as shown in Eq.~\eqref{eq_context_feature}

				\ForEach {\text{temporal distance}  $\Delta t=1,\!\ldots,\!\text{MAX}$ \text{and node} $u\in V$} \label{algline:for-step-distance}
				\State Obtain $u$'s final histogram $\vh(\mathcal{C}^{\deltat}_u)$ over its contexts using        Eq.~\eqref{eq_aggregation}
				\State Obtain a $K^\deltat$-dim, sparse, binary hashcode $z_u^\deltat$ based on (modified) SimHash (\S~\ref{sec:aggr-hashing}) \EndFor

				\State Obtain the binary \textsc{n2b} embeddings $\vz_u$ of all nodes across temporal distances $\deltat$ via Eq.~\eqref{eq_z_representation}
				\State Perform (un)supervised user stitching via binary classification or hashing (\S~\ref{sec:setup},\ref{sec:unsupervised})
				\label{alg-rel-func-learning-framework}
			\end{algorithmic}
		\end{spacing}
	}
\end{algorithm}

Given these representations, we can perform user stitching by casting the problem as supervised binary classification or an unsupervised task based on the output of hashing (Equation~\eqref{eq_z_representation}), which we discuss in \S~\ref{sec:setup}. Putting everything together, we give the pseudocode of \method in Algorithm~\ref{alg_workflow} and its detailed version (for reproducibility) in Appendix~\ref{app:algorithm}.
The runtime computational complexity of \method is $\mathcal{O}(MRL + NK)$, which is linear to the number of edges when $M\gg N$ as $K$ is relatively small (\textbf{R3}). The output space complexity is $\mathcal{O}(NK)$-bit. \method requires even less storage if the binary vectors are represented in the sparse format (see \S~\ref{sec:exp-space-efficiency} for empirical results). 
We provide detailed runtime complexity and space analysis in appendix~\ref{sec:theory}.

\vspace{-0.2cm}
\section{Experiments} 
\label{sec:exp}

We perform extensive experiments on  real-world heterogeneous networks to answer the following questions:
({\bf Q1})~Is \method effective in the user stitching task, and how does it compare to traditional stitching and embedding methods? (\S~\ref{sec:exp-er}-\ref{sec:unsupervised}) ({\bf Q2})~Does \method have low space requirements, and is it more space-efficient than the baselines? (\S~\ref{sec:exp-space-efficiency})
({\bf Q3})~Is \method scalable? (\S~\ref{sec:exp-scalability})

\eat{
	Before we provide answers to these questions, we describe the datasets, our setup for the user stitching task, and the baselines that we use in our experiments. }
\subsection{Experimental Setup}
\label{sec:setup}
We ran our analysis on Mac OS platform with 2.5GHz Intel Core i7, 16GB RAM.

\begin{wraptable}{r}{7cm}
	\vspace{-0.85cm}
	\centering
	\caption{Network statistics and properties for our six real-world datasets. `D’: directed; `W’: weighted; `H’: heterogeneous; `T’: temporal network. }
	\vspace{-0.2cm}
	\centering
	{\scriptsize 
		\setlength{\tabcolsep}{3pt} \begin{tabular}{lrrc  cccc }
			\multicolumn{1}{c}{\textbf{Data}} & {\bf Nodes} & {\bf Edges} & 
			{\bf $|T_V|$} & 
			\rotatebox{0}{\bf D} &
			\rotatebox{0}{\bf W} &
			\rotatebox{0}{\bf H} &
			\rotatebox{0}{\bf T} 
			\\ 
			\noalign{\hrule height 0.8pt}
			citeseer & 4460 & 2892 & 2 & 
			& \checkmark & \checkmark & 
			\\
			yahoo & 100,058 & 1,057,050 & 2 & 
			\checkmark & \checkmark & \checkmark & 
			\\
			bitcoin & 3,783 & 24,186 & 1 & 
			\checkmark & \checkmark & & \checkmark
			\\
			digg & 283,183 & 6,473,708 & 2 & 
			& & \checkmark & \checkmark
			\\
			wiki & 1,140,149  & 7,833,140 & 1 & 
			\checkmark & & & \checkmark
			\\
			comp-X & 5,500,802 & 5,291,270 & 2 &
			\checkmark & \checkmark & \checkmark & \checkmark
			\\
			
			\noalign{\hrule height 0.7pt}
		\end{tabular}
	}
	\label{table_stats_t}
	\vspace{-0.6cm}
\end{wraptable}
\vspace{-0.4cm}
\subsubsection{Data.}
We use five real-world heterogeneous networks from the Network Repository~\cite{nr}, as well as a real, proprietary user stitching dataset, `Company X' web logs. The latter data form a temporal heterogeneous network consisting of web sessions of user devices and their IP addresses.
High degree nodes representing anomalous behavior (\eg, bots or public WiFi hotspots) are filtered out.
Our framework is also capable of modeling domain-specific features, such as user-agent strings and geolocation~\cite{kim2017probabilistic}, if this is available.  Even without them, however, it achieves strong performance.
We give the statistics of all the networks in Table~\ref{table_stats_t}, and additional details in Appendix~\ref{app:data}.

\vspace{-0.4cm}
\subsubsection{Task Setup.}
With the exception of \S~\ref{sec:unsupervised}, we cast the user stitching task as a binary classification problem, where for each pair of nodes we aim to predict whether they correspond to the same entity (\ie, we should stitch them).  We use logistic regression with regularization strength $1.0$ and stopping criterion $10^{-4}$.

For the real user stitching data, we use the held-out, ground-truth information to evaluate our method.
For the five real-world networks without known user pairs, we introduce user replicas following a similar procedure to~\cite{bhattacharya:07}: 
we sample $5\%$ of the nodes with degrees larger than average, introduce a replica $u'$ for each sampled node $u$, and distribute the original edges between $u$ and $u'$.
Specifically, each edge remains connected to $u$ with probability $p_1$, otherwise it connects to the replica node $u'$. Additionally,
each edge that is incident to $u$ has probability $p_2$ to also connect to $u'$. 
Unless otherwise specified, we use $p_1=0.6$ and $p_2=0.3$.

Given the positive replica pairs, we sample an equal number of negative pairs uniformly at random and include these in the training and testing sets.  Comp-X, the dataset with ground-truth replicas, also has pre-defined approximately 50/50 training-testing splits that we use.  
Afterwards, embeddings are derived for each node pair by concatenation: $[\vz(u), \vz(u')]$.
Using these node pair embeddings, we learn a logistic regression (LR) model and use it to predict the node pairs that should be stitched in the held-out test set.
These are the nodes that correspond to the same entity.
We measure the predictive performance of all the methods in terms of AUC, accuracy and $F1$ score.

\vspace{-0.4cm}
\subsubsection{Baselines.}
We compare to various methods that target different network types: \vspace{-0.2cm}
\begin{itemize}
	\item {\bf Homogeneous networks}: \textit{Static}--
	{\bf (1)}~Spectral embeddings or SE~\cite{von2007tutorial},
	{\bf (2)}~LINE~\cite{line},
	{\bf (3)}~DeepWalk or DW~\cite{deepwalk}, 
	{\bf (4)}~node2vec or n2vec~\cite{node2vec}, 
	{\bf (5)}~struc2vec or s2vec~\cite{ribeiro2017struc2vec}, and  
	{\bf (6)}~DNGR~\cite{cao2016deep}. \textit{Temporal}-- {\bf (7)} CTDNE~\cite{nguyen2018continuous}.
	
	\item {\bf Heterogeneous networks}:
	{\bf (8)}~Common neighbors (CN)~\cite{bhattacharya:07},             {\bf (9)}~metapath2vec or m2vec~\cite{dong2017metapath2vec}, and
	{\bf (10)}~AspEm~\cite{shi2018aspem}.
	
\end{itemize}
\vspace{-0.1cm}
The baselines are configured to achieve the best performance, for $K=128$-dimensional embeddings, according to  the respective papers. 
For reproducibility, we describe the settings in Appendix~\ref{app:config}.

\vspace{-0.4cm}
\subsubsection{\method Setup \& Variants.} 
Similar to the baselines, \method performs $R=10$ walks per edge, with length up to $L=20$, and we set the max temporal distance $\text{MAX}=3$. We justify these decisions in Appendix~\ref{sec:sens_analysis}. On the largest dataset, Comp-X, we use a maximum walk length $L=5$ and temporal distance $\text{MAX}=2$.
While various node attributes can be given as input to \method, for consistency we derive and use the total, in-/out-degree of each node in $\mF$. 
We experiment with different variants of \method (or \textsc{n2b} for short): 
(1)~{\bf \methodbase} applies to static networks; 
(2)~{\bf \methodshort} uses the short-term tactic in the random walks (\S~\ref{sec:RW});  
(3)~{\bf \methodlong} uses the long-term tactic;  and
(4)~{\bf \methodunsup} targets unsupervised user stitching, which most baselines cannot handle (except for CN). 
To explore our method's performance in unsupervised settings (\S~\ref{sec:unsupervised}), we directly `cluster' the LSH-based, binary node representations $\vz_{u}$ generated by \methodbase. 
The idea is that nodes that hash to the same `bucket' likely map to the same entity and should be stitched. To map entities to buckets we use the banding technique~\cite{rajaraman2014mining}:
per band---one per representation $\vz^\deltat$ at temporal distance $\deltat$---we apply AND-construction on the output of bit sampling, and then OR-construction across the bands.

\subsection{Accuracy in Supervised User Stitching}
\label{sec:exp-er}
\eat{TODO: Should we use visitor stitching (as opposed to entity resolution)?; visitor stitching is a more specific problem that uses web logs to match devices/web sessions to a single user. We could also have a subsection under entity resolution for visitor stitching on Adobe data.}

We start by evaluating the predictive performance of \method for supervised user stitching on both static and temporal networks.

\vspace{-0.3cm}
\subsubsection{Static Networks.}\label{sec:exp-er-static}

Here we evaluate the effectiveness of multi-dimensional feature contexts.
Since static networks lack temporal information, \method performs random walks similarly to existing works to collect nodes in structural contexts.
The main difference lies in representing diverse feature histograms.
We run \method against both homogeneous and heterogeneous baselines as shown in Table~\ref{table:entity-resolution-static}, and observe that it performs the best in most evaluation metrics on both graphs.
\method outperforms existing random-walk based methods as expected: node IDs in the contexts is distorted by the replicas generated, thus feature-based methods should prevail.
This is also validated by the results for struc2vec, which captures the equivalency of structural feature sequences in embeddings.
metapath2vec and LINE achieve promising result on yahoo but not on citeseer,  
as the latter is an undirected bipartite graph, node distributions of the 2-order contexts explored by LINE are highly correlated and indistinguishable for stitching. 
On the contrary, CN (common-neighbors) yields promising result on citeseer but not yahoo. This is likely due to the graph structure,  
which we explain in more detail in Sec.\ref{sec:unsupervised}. 
We encountered out-of-memory errors for DNGR due to the algorithmic complexity and out-of-time-limit for struc2vec.

\begin{Result} 
	\label{res:er-accuracy-1}
	On static graphs, \method achieves comparable performance in AUC, and slightly better $F$1 score with $0.60\%-2.10\%$ improvement over baselines in the stitching task.
\end{Result}
\vspace{-0.1cm}

\begin{table*}[t!]
	\centering
	\caption{Entity resolution results for \emph{static} networks. Our method outperforms all the baselines. 
		{\scriptsize | OOT = Out Of Time (6h);  OOM = Out Of Memory (16GB). The asterisk $^{\ast}$ denotes statistically significant improvement over the best baseline at $p<0.05$ in a two-sided t-test.}
	}
	\label{table:entity-resolution-static}
	{\scriptsize
		\begin{tabular}{lc H C{0.95cm} C{0.95cm} C{0.95cm} C{1.0cm} C{0.95cm} C{1.07cm} C{0.98cm} C{1.0cm} C{0.98cm} | C{1.0cm} H}
			\toprule
			& Metric & Cosum & CN & SE & LINE & DW & n2vec & s2vec &
			DNGR & m2vec & AspEm & 
			\abbrebase & n2b-st
			\\ \midrule
			\rotatebox[origin=c]{90}{citeseer}
			& \begin{tabular}{@{}r@{}}\textbf{AUC} \\ \textbf{ACC} \\ \eat{\textbf{Precision} \\ \textbf{Recall} \\} 
				\textbf{F1}
			\end{tabular}
			& \begin{tabular}{@{}c@{}} - \\ - \\ \eat{- \\ - \\} - \end{tabular} & \begin{tabular}{@{}c@{}} 0.9141 \\ 0.9141 \\ \eat{ 1.0 \\ 0.8482 \\} 0.9137 \end{tabular} & \begin{tabular}{@{}c@{}} 0.4846 \\ 0.5045 \\ \eat{0.5051 \\ 0.4464 \\} 0.5028 \end{tabular} & \begin{tabular}{@{}c@{}} 0.5481 \\ 0.5372 \\ \eat{0.5361 \\ 0.5532 \\} 0.5371 \end{tabular} & \begin{tabular}{@{}c@{}} 0.5614 \\ 0.5579 \\ \eat{0.5495 \\ 0.6421 \\} 0.5547 \end{tabular} & \begin{tabular}{@{}c@{}} 0.6188 \\ 0.6211 \\ \eat{0.5983 \\ 0.7368 \\} 0.6159 \end{tabular} & \begin{tabular}{@{}c@{}} 0.9344 \\ 0.8936 \\ \eat{0.8304 \\ 0.9894 \\} 0.8926 \end{tabular} & \begin{tabular}{@{}c@{}} 0.5015 \\ 0.4688 \\ \eat{0.4667 \\ 0.4375 \\} 0.4682 \end{tabular} & \begin{tabular}{@{}c@{}} 0.5546 \\ 0.5357 \\ \eat{0.5392 \\ 0.4911 \\} 0.5348 \end{tabular} & \begin{tabular}{@{}c@{}}0.5049 \\ 0.5223 \\ \eat{0.5217 \\ 0.5357 \\} 0.5222 \end{tabular} & \begin{tabular}{@{}c@{}} \cellcolor{gray!25}0.9480$^{\ast}$ \\ \cellcolor{gray!25}0.9196$^{\ast}$ \\ \eat{0.8672 \\ 0.9911 \\} \cellcolor{gray!25}0.9192$^{\ast}$ \end{tabular} \\\midrule
			\rotatebox[origin=c]{90}{yahoo}
			& \begin{tabular}{@{}r@{}}\textbf{AUC} \\ \textbf{ACC} \\ \eat{\textbf{Precision} \\ \textbf{Recall} \\} \textbf{F1}\end{tabular}
			& OOM & \begin{tabular}{@{}c@{}} 0.6851 \\ 0.6851 \\ \eat{1.0 \\ 0.3703 \\} 0.6505 \end{tabular} & \begin{tabular}{@{}c@{}} 0.5378 \\ 0.4760 \\ \eat{0.4497 \\ 0.2143 \\} 0.4375 \end{tabular} & \begin{tabular}{@{}c@{}} 0.8050 \\ 0.7771 \\ \eat{0.7500 \\ 0.8313 \\} 0.7764 \end{tabular} & \begin{tabular}{@{}c@{}} 0.7640 \\ 0.7117 \\ \eat{0.7063 \\ 0.72491 \\} 0.7117 \end{tabular} & \begin{tabular}{@{}c@{}} 0.7636 \\ 0.7233 \\ \eat{0.7126 \\ 0.7485 \\} 0.7231 \end{tabular} & OOT & OOM & \begin{tabular}{@{}c@{}} \cellcolor{gray!25}0.8233 \\ 0.7827 \\ \eat{0.7126 \\ 0.7485 \\} 0.7823 \end{tabular} & \begin{tabular}{@{}c@{}} 0.4938 \\ 0.5018 \\ \eat{0.5018 \\ 0.5030 \\} 0.5018 \end{tabular} & \begin{tabular}{@{}c@{}} 0.8088 \\ \cellcolor{gray!25}\cellcolor{gray!25}0.8010 \\ \eat{0.7481 \\ 0.9076 \\} \cellcolor{gray!25}0.7987 \end{tabular} 
			
			\\
			\bottomrule
			
		\end{tabular}
	}
	\vspace{-0.4cm}
\end{table*}

\vspace{-0.6cm}
\subsubsection{Temporal Networks.}\label{sec:exp-er-temporal}
Table ~\ref{table:entity-resolution-dynamic} depicts the stitching performance of \method using both the short- and long-term tactics against the same set of baselines used in static graphs as well as CTDNE, an embedding framework designed for temporal graphs.
We exclude metapath2vec, as metapaths are not meaningful in homogeneous networks, and the method ran out of time for the heterogeneous networks.
We observe that \methodshort outperforms \methodlong in most cases, which is reasonable because \methodlong derives shorter contexts constrained by temporal-order.
We also justify the effectiveness of temporal random walk by comparing it with both \methodbase and static baselines where we only make use of the graph structures without specifying edge timestamps.
We observe that \methodbase is the best-performing method for the digg dataset and Comp-X over the temporal variants of \method.
The reason behind this is that there is a tradeoff in constraining temporal walks to respect time: we more accurately model realistic sequences of events at the cost of restricting the possible context. On these particular temporal graphs, walks may already be limited in length by the bipartite structure, so the latter cost becomes more appreciable.  Nevertheless, both static and dynamic versions of \method almost always outperform other baselines.
In particular, across all datasets, \methodshort still outperforms the temporal baseline, CTDNE in all cases, which further demonstrates the effectiveness of multi-feature aggregation.

\begin{table*}[t!]
	{\scriptsize
		\centering
		\caption{Entity resolution results for \emph{temporal} networks: strong performance for \method variants.
			{\scriptsize | OOT = Out Of Time (6h);  OOM = Out Of Memory (16GB); $^{\ast}$ denotes statistically significant improvement over the best baseline at $p<0.05$ in a two-sided t-test.}
		}
		\label{table:entity-resolution-dynamic}
		\resizebox{1.0\textwidth}{!}{
			\fontsize{8.5}{9.5}\selectfont
			\begin{tabular}{lc H C{0.95cm} C{0.95cm} C{0.95cm} C{1.0cm} C{0.95cm} C{1.0cm} C{1.07cm} C{0.98cm} H C{1.1cm} | C{1.1cm} C{1.1cm} C{1.1cm}}
				\toprule
				& Metric & Cosum & CN &  SE & LINE & DW & n2vec & s2vec &
				DNGR & AspEm & m2vec & CTDNE &
				\abbrebase & \abbreshort & \abbrelong 
				\\ \midrule
				\rotatebox[origin=c]{90}{bitcoin}
				& \begin{tabular}{@{}r@{}}\textbf{AUC} \\ \textbf{ACC} \\ \eat{\textbf{Precision} \\ \textbf{Recall} \\} \textbf{F1}\end{tabular}
				& \begin{tabular}{@{}c@{}} - \\ - \\ \eat{- \\ - \\} - \end{tabular} & \begin{tabular}{@{}c@{}} 0.7474 \\ 0.7174 \\ \eat{1.0 \\ 0.4947 \\} 0.7001 \end{tabular} & \begin{tabular}{@{}c@{}} 0.5828 \\ 0.5842 \\ \eat{0.6250 \\ 0.4211 \\} 0.5728 \end{tabular} & \begin{tabular}{@{}c@{}} 0.6071 \\ 0.5842 \\ \eat{0.5755 \\ 0.6421 \\} 0.5828 \end{tabular} & \begin{tabular}{@{}c@{}} 0.6306 \\ 0.6158 \\ \eat{0.6146 \\ 0.6211 \\} 0.6158 \end{tabular} & \begin{tabular}{@{}c@{}} 0.6462 \\ 0.6158 \\ \eat{0.6122 \\ 0.6316 \\} 0.6157 \end{tabular} & \begin{tabular}{@{}c@{}} \cellcolor{gray!25} 0.8025 \\ 0.7263 \\ \eat{0.7263 \\ 0.7263 \\} 0.7263 \end{tabular} & \begin{tabular}{@{}c@{}} 0.5909 \\ 0.5526 \\ \eat{0.5510 \\ 0.5684 \\} 0.5525 \end{tabular} & \begin{tabular}{@{}c@{}} 0.5344 \\ 0.5316 \\ \eat{ 0.5326 \\ 0.5158 \\} 0.5315 \end{tabular} & \begin{tabular}{@{}c@{}} - \\ - \\ \eat{- \\ - \\} - \end{tabular} & \begin{tabular}{@{}c@{}} 0.6987 \\ 0.6000 \\ \eat{0.6234 \\ 0.5053 \\} 0.5964 \end{tabular} & \begin{tabular}{@{}c@{}} 0.7584 \\ 0.7211 \\ \eat{0.7100 \\ 0.7474 \\} 0.7209 \end{tabular} & \begin{tabular}{@{}c@{}} 0.7609 \\ \cellcolor{gray!25}0.7268 \\ \eat{0.7647 \\ 0.6842 \\} \cellcolor{gray!25}0.7271 \end{tabular} & \begin{tabular}{@{}c@{}} 0.7380 \\ 0.6737 \\ \eat{0.6667 \\ 0.6947 \\} 0.6735 \end{tabular} \\\midrule

				\rotatebox[origin=c]{90}{digg}
				& \begin{tabular}{@{}r@{}}\textbf{AUC} \\ \textbf{ACC} \\ \eat{\textbf{Precision} \\ \textbf{Recall} \\} \textbf{F1}\end{tabular}
				& OOM & \begin{tabular}{@{}c@{}} 0.6217 \\ 0.6217 \\ \eat{1.0 \\ 0.2434 \\} 0.5585 \end{tabular} & \begin{tabular}{@{}c@{}} 0.5171 \\ 0.5152 \\ \eat{0.5564 \\ 0.0516 \\} 0.3770 \end{tabular} & \begin{tabular}{@{}c@{}} 0.7878 \\ 0.7694 \\ \eat{0.7371\\ 0.8376 \\} 0.7683 \end{tabular} & \begin{tabular}{@{}c@{}} 0.7398 \\ 0.6971 \\ \eat{ 0.6763 \\ 0.7562 \\} 0.6960 \end{tabular} & \begin{tabular}{@{}c@{}} 0.7445 \\ 0.7013 \\ \eat{0.6809 \\ 0.7576 \\} 0.7003 \end{tabular} & OOT
				& OOM
				& \begin{tabular}{@{}c@{}} 0.5105 \\ 0.5088 \\ \eat{0.5087 \\ 0.5138 \\} 0.5088 \end{tabular} & \begin{tabular}{@{}c@{}} - \\ - \\ \eat{- \\ - \\} - \end{tabular} & \begin{tabular}{@{}c@{}} 0.6967 \\ 0.5915 \\ \eat{0.6110 \\ 0.5058 \\} 0.5884 \end{tabular} & \begin{tabular}{@{}c@{}} \cellcolor{gray!25}0.8185$^{\ast}$ \\ \cellcolor{gray!25}0.7982$^{\ast}$ \\ \eat{0.7453 \\ 0.9060 \\} \cellcolor{gray!25}0.7958$^{\ast}$ \end{tabular} & \begin{tabular}{@{}c@{}} 0.7611 \\ 0.7418 \\ \eat{ 0.7185 \\ 0.7952 \\} 0.7411 \end{tabular} & \begin{tabular}{@{}c@{}} 0.7587 \\ 0.7444 \\ \eat{0.7171 \\ 0.8071 \\} 0.7433 \end{tabular} \\
				\midrule
				\rotatebox[origin=c]{90}{wiki}
				& \begin{tabular}{@{}r@{}}\textbf{AUC} \\ \textbf{ACC} \\ \eat{\textbf{Precision} \\ \textbf{Recall} \\} \textbf{F1}\end{tabular}
				& OOM & \begin{tabular}{@{}c@{}} 0.6997 \\  0.6997 \\ \eat{1.0 \\ 0.3994 \\} 0.6699 \end{tabular} & OOT & \begin{tabular}{@{}c@{}} 0.7854 \\  0.7132 \\ \eat{0.7274 \\ 0.6819 \\} 0.7129 \end{tabular} & OOM & OOM & OOT
				& OOM
				& \begin{tabular}{@{}c@{}}0.5374 \\ 0.5141 \\ \eat{0.5011 \\ 0.4993 \\} 0.5141 \end{tabular} & \begin{tabular}{@{}c@{}} - \\ - \\ \eat{- \\ - \\} - \end{tabular} & \begin{tabular}{@{}c@{}} 0.7707 \\ 0.6488 \\ \eat{0.7174 \\ 0.4910 \\} 0.6398 \end{tabular} & \begin{tabular}{@{}c@{}} 0.8230 \\ 0.7145 \\ \eat{0.7972 \\ 0.5753 \\} 0.7088 \end{tabular} & \begin{tabular}{@{}c@{}} \cellcolor{gray!25}0.8259$^{\ast}$ \\ \cellcolor{gray!25}0.7510$^{\ast}$ \\ \eat{0.8268 \\ 0.6349 \\} \cellcolor{gray!25}0.7476$^{\ast}$ \end{tabular} & \begin{tabular}{@{}c@{}} 0.8214 \\ 0.7103 \\ \eat{- \\ - \\} 0.7067 \end{tabular} \\\midrule
				
				\rotatebox[origin=c]{90}{comp-X}
				& \begin{tabular}{@{}r@{}}\textbf{AUC} \\ \textbf{ACC} \\ \eat{\textbf{Precision} \\ \textbf{Recall} \\} \textbf{F1}\end{tabular}
				& - & \begin{tabular}{@{}c@{}} 0.5970 \\ 0.5970 \\ \eat{- \\ - \\} 0.5189 \end{tabular} & OOM & \begin{tabular}{@{}c@{}} 0.5000 \\ 0.6757 \\ \eat{- \\ - \\} 0.4032 \end{tabular} & OOM & OOM & OOT & OOM
				& \begin{tabular}{@{}c@{}} 0.5213 \\ 0.5103 \\ \eat{- \\ - \\} 0.5103 \end{tabular} & \begin{tabular}{@{}c@{}} - \\ - \\ \eat{- \\ - \\} - \end{tabular} & OOM & \begin{tabular}{@{}c@{}} \cellcolor{gray!25}0.8095$^{\ast}$ \\ \cellcolor{gray!25}0.8414$^{\ast}$ \\ 
					\cellcolor{gray!25}0.8154$^{\ast}$ \end{tabular}  & \begin{tabular}{@{}c@{}} 0.7496 \\ 0.7959 \\ \eat{- \\ - \\} 0.7581 \end{tabular}  & \begin{tabular}{@{}c@{}} 0.7525 \\ 0.7975 \\ \eat{- \\ - \\} 0.7606 \end{tabular} \\
				
				\bottomrule
			\end{tabular}
		}
		\vspace{-0.5cm}
	}
\end{table*}

\method variants outperform the static methods in nearly all cases except the bitcoin dataset where \methodshort achieves lower AUC than struc2vec but higher ACC and $F$1-score. This is because \method loses some information when representing the node contexts as binary vectors comparing with real-value representation. However, we consider this loss mild as \method still outperforms all the other static baselines. In addition, struc2vec ran out of time on the larger datasets while \method achieves promising performance efficiently with $3.90\%-5.16\%$ improvement in AUC and $3.58\%-4.87\%$ improvement in $F1$ score than the best baseline method.
At the same time, our approach uses much less information than the static methods, since the length of the temporal walks are typically shorter than random walks that do not have to respect time.  

\begin{Result} \label{res:er-accuracy-2}
	Dynamic and static variants of \method outperform the other baselines by up to $5.2\%$ in AUC and $4.9\%$ in $F1$ score. Between the two dynamic variants, the short-term tactic performs better than the long-term one.
\end{Result}
\vspace{-0.6cm}

\vspace{-0.2cm}
\subsubsection{Restricting the Output Space Requirements.} 

To evaluate the performance of stitching with explicit storage requirement, we hash the real-value embeddings given by baselines into binary and achieve output storage to be consistent with \method. We observe that \method still achieves the best performance (refer to Table~\ref{table:justify_hashing} in Appendix~\ref{sec:justify-hashing} for more details).

\subsection{Accuracy in Unsupervised User Stitching}
\label{sec:unsupervised}

As mentioned in \S~\ref{sec:setup}, \method can naturally perform unsupervised user stitching by leveraging the generated node representations as hashcodes. Only nodes mapped to the same `buckets' are candidates  for stitching together.
This process allows us to stitch entities without involving quadratic comparisons between all pairs of nodes in the graph.
Similarly, CN outputs a set of nodes sharing a certain amount of neighbors as the candidates to be stitched together.
We evaluate the quality of hashing given by \methodunsup against CN, and make use of the candidates to predict the testing set of node pairs given by following the same setup in \S~\ref{sec:exp-er} in an unsupervised scheme.

Based on the results in Table~\ref{table:unsupervised}, we observe that \methodunsup outperforms CN on every dataset other than citeseer. 
The reason is  that in this ``author contributes to paper'' dataset, author references appearing in the same set of papers have high probability to correspond to the same researcher in reality.
Therefore the assumption made by CN suits well  this scenario, whereas \method hashes nodes with similar features in the context instead of those with similar neighbor identities (IDs).
For datasets with less strict cross-type relationship, \method achieves $2.81\%-15.12\%$ improvement in accuracy ACC and $4.96\%-26.66\%$ improvement in $F1$ score (including digg, another bipartite graph with inner connected components of the same node types).

\begin{Result} \label{res:unsupervised-stitching-1}
	The unsupervised variant of \method, \methodunsup, outperforms CN on most graphs.
	\eat{For unsupervised stitching, \method outperforms CN on all but one graph (which is the only k-partite graph).
}\end{Result}
\vspace{-0.5cm}

\begin{table}[t!]
	\centering
	{\scriptsize
		\caption{
			Unsupervised stitching performance between CN and \method
		}
		\vspace{1mm}
		\label{table:unsupervised}
		\setlength{\tabcolsep}{3pt} \begin{tabular}{@{}H r
				@{\hskip 0.35cm}ccl@{\hskip 0.35cm} cc l@{\hskip 0.35cm} cc l@{\hskip 0.35cm} cc l@{\hskip 0.35cm} cc }
			\toprule
			\multirow{2}{*}{\bf Method } & \multirow{2}{*}{\bf Metric} & \multicolumn{2}{c}{\bf  citeseer } && \multicolumn{2}{c}{\bf yahoo} && \multicolumn{2}{c}{\bf bitcoin} && \multicolumn{2}{c}{\bf digg} && \multicolumn{2}{c}{\bf wiki}  \\ 
			\cline{3-4} \cline{6-7} \cline{9-10} \cline{12-13} \cline{15-16}
			
			&\TT &   CN & \abbreunsup   &&   CN & \abbreunsup   &&    CN & \abbreunsup   &&   CN & \abbreunsup    &&  CN  & \abbreunsup   \\   
			\midrule

			\multirow{4}{*}{{\bf CN}} & ACC       & \cellcolor{gray!25}0.9141 & 0.8661   &&   0.6851 & \cellcolor{gray!25}0.7553   && 0.7474 & \cellcolor{gray!25}0.7684    &&  0.6217 & \cellcolor{gray!25}0.7157   &&   0.6997 & \cellcolor{gray!25}0.7350 \\
			& F1        & \cellcolor{gray!25}0.9137 & 0.8660   &&   0.6505 & \cellcolor{gray!25}0.7518   && 0.7301 & \cellcolor{gray!25}0.7663    && 0.5585 & \cellcolor{gray!25}0.7074    && 0.6699 & \cellcolor{gray!25}0.7349 \\
			\bottomrule
		\end{tabular}
	}
	\vspace{-0.3cm}
\end{table}

\subsection{Output Storage Efficiency} \label{sec:exp-space-efficiency}
Next we evaluate space efficiency of our proposed method over baselines that output node embeddings. 
Instead of real-value matrices, the binary hashcodes generated by \method can be stored in the sparse format so presumably it should take trivial storage. 
We visualize the storage requirements in Figure~\ref{fig_space} and provide detailed storage usage in Table~\ref{table:space} in Appendix~\ref{sec:detailed_output_space}. 
\begin{Result} \label{res:space-1}
	Compared to the other methods, \method uses between $63\times$ and $339\times$ less space (while always achieving comparable or better stitching performance as shown in \S~\ref{sec:exp-er}).
\end{Result}
\vspace{-0.5cm}

\eat{
	\subsection{Runtime performance of Streaming}

	\begin{table*}[t!]
		{\small
			\centering
			\caption{\rr{TODO: Report average runtime per edge in the stream (should be milliseconds). Do not split the stream. Use the full stream. Also, report number of updates (or nodes updated) per edge in the stream, and any other statistics}
				Runtime of \method on different proportion of input streaming data. We also record the time of \method running on the complete dataset as reference - it is NOT for comparison.
				\dijin{update per edge, average time}
			}
			\label{table_link_prediction}
			\begin{tabular}{lr C{1.25cm} C{1.25cm} C{1.25cm} C{1.25cm} C{1.25cm} C{1.25cm} C{1.25cm} C{1.25cm} C{1.25cm} | C{2.0cm} H}
				\toprule
				\textbf{Data} & $5\%$ & $10\%$ & $20\%$ & $30\%$ & $40\%$ & $50\%$ & complete
				\\ \midrule
				{\bf bitcoin} & 5.13  & 5.63  & 7.31  & 9.82  & 13.11  & 16.56  & 39.97 \\
				{\bf digg} & - & - & - & - & - & - & 3062.13 \\
				\bottomrule
			\end{tabular}
			
		}
	\end{table*}
	
}

\begin{figure*}[!t]
	\centering
	\includegraphics[width=.96\linewidth]{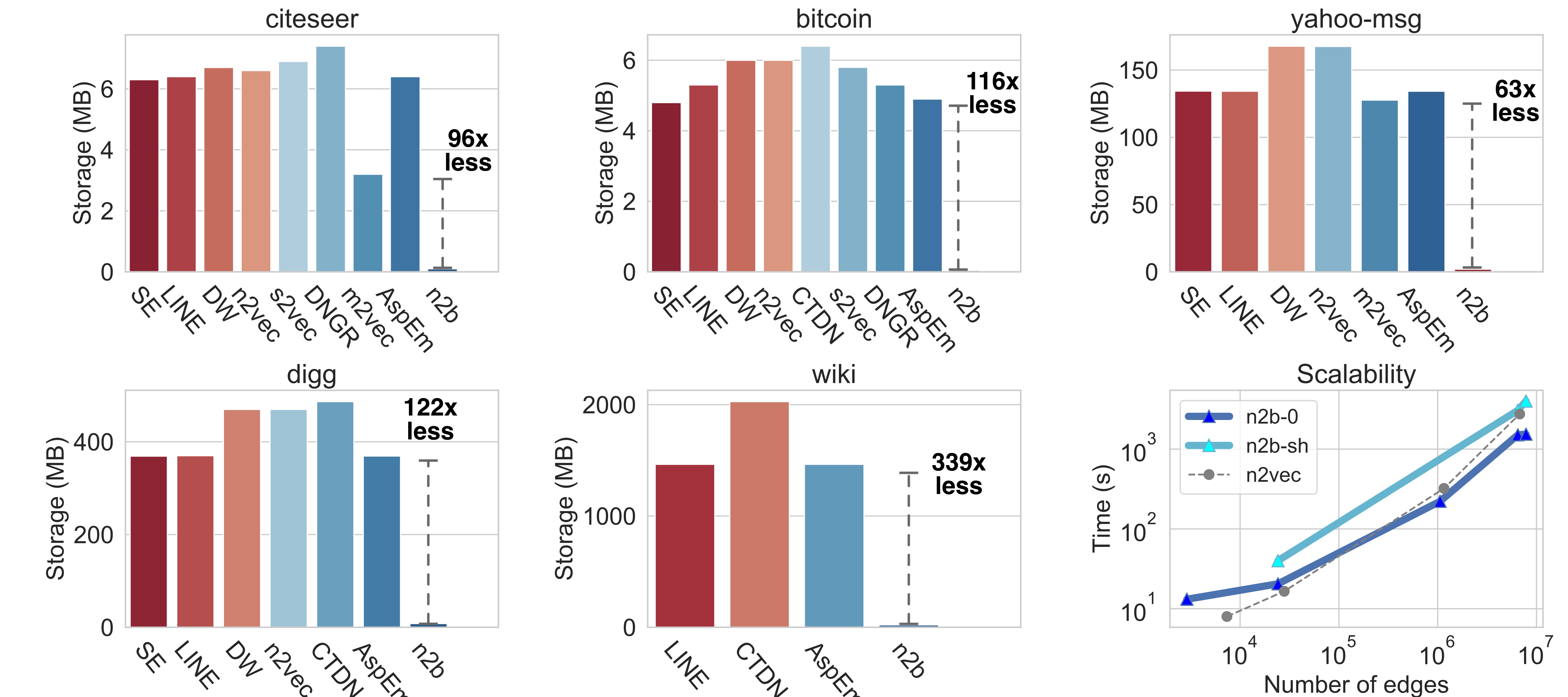}
	\vspace{-0.3cm}
	\caption{First 5 plots: output storage in MB for all the methods that completed successfully in five datasets.
		\method is also shown to be scalable for large graphs.     }
	\label{fig_space}
	\vspace{-0.5 cm}
\end{figure*}

\subsection{Scalability} \label{sec:exp-scalability}

To evaluate the scalability, we report the runtime of applying \method to obtain node representations for the datasets shown in Table~\ref{table_stats_t} versus their numbers of edges. 
We note that \methodshort runs only on temporal networks, i.e., a subset of the datasets. We also visualize the runtime of node2vec as reference, as it is designed for large graphs and is implemented in the same language (Python).
Based on the last subplot in Figure~\ref{fig_space}, we observe that \method scales similarly as node2vec with less runtime space as node2vec ran out of memory on the largest dataset (wiki). 
As shown in Appendix~\ref{sec:theory}, the worst-case time complexity is linear in the edges. 
We give the exact runtimes in Table~\ref{table:runtime-perf} in the Appendix~\ref{sec:detailed_output_space}.

\vspace{-0.2cm}
\section{Related Work} 
\vspace{-0.2cm}
\label{sec:related-work}

\noindent \textbf{Entity Resolution}  (the general problem under which user stitching falls) has been widely studied and applied in different domains such as databases and information retrieval~\cite{dong2009data,getoor2013entity}.
Traditional methods that are based on distances can be broadly categorized into (1)~pairwise-ER~\cite{cohen2002learning}, which independently decide which pairs are same entity based on a distance threshold, and (2)~clustering~\cite{dasgupta2012overcoming}, which links nodes in the same cluster.  
However, these methods are computationally expensive and do not scale to large datasets.
Other techniques range from supervised classification~\cite{saha2015probabilistic} to probabilistic soft logic~\cite{kim2017probabilistic} or fingerprinting~\cite{eckersley2010unique} using side information (\eg, user-agent strings, other web browser features, geo-location).  
These methods tend to be problem- or even data-specific. On the contrary, our method is general by modeling the data with a heterogeneous, dynamic network that uses both node features (optional) \textit{and} graph structure.

\vspace{0.1cm}
\noindent \textbf{Node embeddings} aim to preserve a notion of node similarity into low-dimensional vector space. 
Most general methods~\cite{node2vec,deepwalk,line} and the state-of-the-art for heterogeneous or dynamic networks ~\cite{dong2017metapath2vec,nguyen2018continuous}, define node similarity based on co-occurrence or proximity in the original network (Appendix~\ref{app:related}). However, in the user stitching problem, it is possible that corresponding entities may \eat{exhibit similar behavior (resulting in similar local topologies and features) but} not connect to the same entities, resulting in lower proximity-based similarity. 
Embeddings preserving \textit{structural identity}~\cite{ribeiro2017struc2vec,DonnatZHL18graphwave,xnetmf,JinRKKRK19multilens,JinHSWWSK19ember,LiuZSGK19,role2vec} overcome this drawback. \method additionally handles various graph settings (heterogeneous, dynamic) at greater space efficiency thanks to hashing.

\smallskip
\noindent\textbf{Locality sensitivity hashing (LSH)} was first introduced as a randomized hashing framework for efficient approximate nearest neighbor search in high dimensional space~\cite{indyk1998approximate}.
It specifies a family of hash functions, $\mathcal{H}$, that maps similar items to the same bucket identified through hash codes with higher probability than dissimilar items~\cite{rajaraman2014mining}. LSH families for different distances have been widely studied, such as SimHash for cosine distance~\cite{charikar2002similarity}, min-hash for Jaccard similarity~\cite{broder1997syntactic}, 
and more (Appendix~\ref{app:related}). 
In our work, we leverage LSH to construct similarity-preserving and space-efficient node representations  for user stitching.

\vspace{-0.2cm}
\section{Conclusion} 
\vspace{-0.2cm}
\label{sec:conc}
We have proposed a hash-based network representation learning framework for identity stitching called \method.
It is both time- and attribute-aware, while also deriving space-efficient sparse binary embeddings of nodes in large temporal heterogeneous networks.
\method uses the notion of feature-based temporal walks to capture the temporal and feature-based information in the data.
Feature-based temporal walks are a generalization of walks that obey time while also incorporating features (as opposed to node IDs).
Using these walks, \method generates contexts/sequences of temporally valid feature values.
Experiments on real-world networks demonstrate the utility of \method as it outputs space-efficient embeddings that use orders of magnitude less space compared to the baseline methods while achieving better performance in user stitching.
An important practical consideration in the application of our work to user stitching is the balance of greater personalization with user privacy.

\vspace{-0.3cm}
\section*{Acknowledgements}
\vspace{-0.2cm}
{
\footnotesize 
This material is based upon work supported by the National Science Foundation under Grant No. IIS 1845491, Army Young Investigator Award No. W911NF1810397, an Adobe Digital Experience and an Amazon research faculty award.
Any opinions, findings, and conclusions or recommendations expressed in this material are those of the author(s) and do not necessarily reflect the views of the NSF or other funding parties. }

\vspace{-0.2cm}
\bibliographystyle{splncs04}
\footnotesize{\bibliography{paper}}

\newpage
\appendix
{\center \textbf{\LARGE Appendix}}
\section{Detailed Algorithm} \label{app:algorithm}

In \S~\ref{sec:approach} we gave the overview of our proposed method, \method. For reproducibility, here we also provide its more detailed pseudocode.

\vspace{-0.3cm}
\definecolor{lightgray}{rgb}{0,0.08,0.45} 
\algblockdefx[parallel]{ParFor}{EndPar}[1][]{$\textbf{parallel for}$ #1 $\textbf{do}$}{$\textbf{end parallel}$}
\algrenewcommand{\alglinenumber}[1]{\fontsize{6.5}{7}\selectfont#1}
\algtext*{EndPar}

\algblockdefx[parallel]{parfor}{endpar}[1][]{$\textbf{parallel for}$ #1 $\textbf{do}$}{$\textbf{end parallel}$}
\algrenewcommand{\alglinenumber}[1]{\scriptsize#1:}

\algblockdefx[foreach1]{foreach}{endforeach}[1][]{$\textbf{for}$ #1 $\textbf{do}$}{$\textbf{end for}$}
\algrenewcommand{\alglinenumber}[1]{\scriptsize#1:}
\begin{algorithm}[h!]
	\caption{\,\small \fontsize{9}{10.5}\selectfont
		\method framework in detail }
	\label{alg_workflow_appendix}
	{\begin{spacing}{1.15}
			\fontsize{7}{8}\selectfont
			\begin{algorithmic}[1]
				\fontsize{8}{9}\selectfont
				\vspace{0.5mm}
				\Require ~\newline
				(un)directed heterogeneous graph $G(V, E, \psi,\xi)$,
				\newline
				number of random walks $R$ per edge, 
				max walk length $L$,
				\newline
				max temporal distance $MAX$, \newline
				embedding dimensionality $K^\deltat$ at temporal dist.\ $\deltat$ (output dim.\  $K =   \sum^{MAX}_{\deltat=1}K^\deltat$)
				\eat{\newline\quad\quad 
					the maximum number of layers to learn $t_{\max}$
				}\medskip
				\fontsize{8}{9}\selectfont
				
				\State Construct $N \times |\mathcal{F}|$ feature matrix $\mF$\textcolor{lightgray}{\Comment{\textsf{Matrix with node attributes and derived features}}}
				\State $\mathcal{C}^{\deltat}_u \leftarrow \emptyset,  \; \forall u\in V, \; \deltat \in [1, \ldots, MAX]$  \textcolor{lightgray}{\Comment{\textsf{Context of $u$: nodes at  temporal distance $\deltat$}}}
				\label{algline-app:typed-base-functions}
				\vspace{1.0mm}
				
				\For {edge $e$ and $\text{ walk} = 1, \! \ldots, \!R $} \label{algline-app:for-walks}
				\textcolor{lightgray}{\Comment{\textsf{Perform $R$ feature-based temporal rand.\ walks per edge}}}

				\State $\langle w_L\rangle_{L\in\mathbb{N}} \leftarrow$ up to $L$-step temporal walk starting from edge $e$ based on tactic                                 \textcolor{lightgray}{\Comment{\textsf{Dfn.~\ref{def:temporal-walk}}}}
				\State $\mathcal{C}^{\deltat}_u \leftarrow$  update the context of nodes $u \in \langle w_L\rangle_{L\in\mathbb{N}}$  and all temporal distances $\deltat$                         
				
				\For {$j=1, \ldots,|\mathcal{F}|$} \textcolor{lightgray}{\Comment{\textsf{Iterate over all the features in matrix $\mF$}}}      \State \textcolor{lightgray}{\Comment{\textsf{Generate the feature-based context by replacing $v \in \mathcal{C}^{\deltat}_u$ with the $f_{v,j}$. $\quad\quad\quad\quad$}}}
				\State \textcolor{lightgray}{\Comment{\textsf{Equivalent to context generation after a feature-based temporal walk (Dfn.~\ref{def:attr-temporal-random-walk}).$\quad\,$}}}
				
				\State $\mathcal{C}^{\deltat}_u|f_{(j)} \leftarrow$  update the feature-based context of  $u\in \langle w_{L}\rangle_{L\in\mathbb{N}}$         \EndFor
				\EndFor
				\State
				\For {$\Delta t=1,\!\ldots,\!MAX$} \label{algline-app:for-step-distance}
				\State Generate $K^\deltat$ random hyperplanes     \ForEach {$\text{node } u\in V$} \textcolor{lightgray}{\Comment{\textsf{For each node, summarize its context with}}}         \ForEach {$\text{node } v\in \mathcal{C}^{\deltat}_u$} \textcolor{lightgray}{\Comment{\textsf{a histogram per feature and node type.}}}

				\State $\vh(\mathcal{C}^{\deltat}_u)=\text{concatenate}[\vh(\mathcal{C}^{\deltat}_u\;|\;f_{(1)}, \nodetype_1),\ldots,\vh(\mathcal{C}^{\deltat}_u\;|\;f_{(|\mathcal{F}|)}, \nodetype_{|\mathcal{T}_V|})]$ \textcolor{lightgray}{\Comment{(Eq.~\eqref{eq_context_feature})}}
				
				\EndFor
				
				\State \textcolor{lightgray}{\Comment{\textsf{Obtain a sparse, binary hashcode per node based on (modified) SimHash. $\quad\;\;$}}}
				\State $z_u^\deltat \leftarrow$ SimHash of node histogram $\vh(\mathcal{C}^{\deltat}_u)$ at distance $\deltat$
				\EndFor
				\State $\mathbf{Z}^\deltat \leftarrow$ $N \times K^\deltat$ matrix with each node's SimHash code per row
				\EndFor

				\State {\bf return} Sparse node representation  $\mathbf{Z}=\text{concatenate}[\mathbf{Z}^1, \ldots, \mathbf{Z}^{MAX}]$ 
				\label{alg-rel-func-learning-framework-detailed}
			\end{algorithmic}
		\end{spacing}
	}
\end{algorithm}

\vspace{-0.45cm}
\section{Complexity Analysis}  
\label{sec:theory}

\textbf{Time Complexity.}
The runtime complexity of \method includes deriving 
(1)~the set of $R$ temporal random walks of length up to $L$, which is $\mathcal{O}(MRL)$ in the worst case; (2)~the feature values of nodes in the walks from step (1); and 
(3)~hashing the feature values of nodes in the context through random projection, which is $\mathcal{O}(NK)$. 
Thus, the total runtime complexity is $\mathcal{O}(MRL + NK)$, which is linear to the number of edges when $M\gg N$ as $K$ is relatively small (\textbf{R3}).
\vspace{0.1cm}

\noindent \textbf{Runtime Space Complexity.}
The space required in the runtime consists three parts: 
(1) the set of temporal random walks (represented as vectors) per edge with complexity $\mathcal{O}(MRL)$, 
(2) the histograms of feature contexts $ N|\mathcal{F}||\mathcal{T}_V|$, and 
(3) the set of randomly-generated hyperplanes $NK$. Therefore, the total runtime space complexity is $\mathcal{O}(MRL + N(|\mathcal{F}||\mathcal{T}_V|+K))$. 
\vspace{0.1cm}

\noindent \textbf{Output Space Complexity.}
The output space complexity of \method is $\mathcal{O}(NK)$-bit. The space required to store binary vectors is guaranteed to be $32\times$ less than vectors represented with real-value floats (4 bytes) with the same dimension. In practice, \method requires even less storage if the binary vectors are represented in the sparse format (see Section~\ref{sec:exp-space-efficiency} for empirical results).

\section{Data Description}
\label{app:data}

Below we provide a more detailed description of the network datasets that we use in our experiments (Table~\ref{table_stats_t}).
\begin{itemize}    
	
	\item \textbf{citeseer}: {CiteSeerX} is an undirected, heterogeneous network that contains the bipartite relations between authors and papers they contributed.
	
	\item \textbf{yahoo}: {Yahoo! Messenger Logs} is a heterogeneous network capturing message exchanges between users at different locations (node attribute).
	
	\item \textbf{bitcoin}: soc-bitcoinA is a who-trusts-whom network on the Bitcoin Alpha platform. The directed edges indicate user ratings.     
	\item \textbf{digg}: This heterogeneous network consists of users voting stories that they like and forming friendships with other users. 
	
	\item \textbf{wiki}: {wiki-talk} is a temporal homogeneous network capturing Wikipedia users editing each other's Talk page over time.

	\item \textbf{comp-X}: A temporal heterogeneous network is derived from a company's  web logs and consists of web sessions of users and their IPs. 
	In the stitching task, we predict the web session IDs that correspond to the same user. 
\end{itemize}

\section{Configuration of Baselines}
\label{app:config}

As we mentioned in \S~\ref{sec:setup}, we configured all the baselines to achieve the best performance according to  the respective papers. 
For all the baselines that are based on random walks (\ie, node2vec, struc2vec, DeepWalk, metapath2vec, CTDNE), we set the number of walks to 20 and the maximum walk length to $L=20$.
For node2vec, we perform grid search over $p,q \in \{0.25, 0.50, 1, 2, 4\}$ as mentioned in~\cite{node2vec} and report the best performance. 
For metapath2vec, we adopt the recommended meta-path ``Type 1-Type 2-Type 1'' (\eg, type 1 = author; type 2 = publication). In DNGR, we set the random surfing probability $\alpha=0.98$ and use a 3-layer neural network model where the hidden layer has 1024 nodes.
We use 2nd-LINE to incorporate 2nd-order proximity in the graph.
For all the embedding methods, we set the embedding dimension to $K=128$. 
Unlike those, CN outputs clusters, each of which corresponds to one entity.

\section{Additional Empirical Analysis}

\subsection{Justification of hashing} \label{sec:justify-hashing}
In this experiment we hash the outputs given by baseline embedding methods using SimHash~\cite{charikar2002similarity} and then perform stitching on two temporal graphs to study their performance under the constraint of storage comparable to \method.
Based on Table~\ref{table:justify_hashing}, we observe fluctuation in the stitching performance of baseline methods, for example, almost all baselines got degenerated scores in all metrics on the bitcoin dataset, especially for struc2vec. On the other hand, however, node2vec, LINE and CTDNE got slight increased scores on yahoo dataset, which is likely due to the fact that the small real-values are amplified when hashed into binary for logistic regression binary classification. It is also possibly due to the graph strucutre. We leave furhter discussion in the future work, but nevertheless, \method outperforms these baselines in all cases. 
This empirical experiment demonstrates that \method effectively preserves context information in the binary hashcodes.

\begin{table*}[t!]
	\vspace{-0.4cm}
	\centering
	\caption{Justification of hashing
	}
	\label{table:justify_hashing}
	\small
	\fontsize{8.5}{9.5}\selectfont
	\begin{tabular}{lc H H C{0.95cm} C{0.95cm} C{1.0cm} C{0.95cm} C{1.1cm} C{1.07cm} H H C{1.04cm} | C{0.92cm} C{0.92cm} C{0.92cm}}
		\toprule
		& Metric & Cosum & CN &  SC$^{*}$ & LINE$^{*}$ & DW$^{*}$ & n2vec$^{*}$ & CTDNE & s2vec$^{*}$ &
		DNGR$^{*}$ & m2vec & AspEm &
		\abbrebase & \abbreshort & \abbrelong
		\\ \midrule
		\rotatebox[origin=c]{90}{bitcoin}
		& \begin{tabular}{@{}r@{}}\textbf{AUC} \\ \textbf{ACC} \\ \eat{\textbf{Precision} \\ \textbf{Recall} \\} \textbf{F1}\end{tabular}
		& \begin{tabular}{@{}c@{}} - \\ - \\ \eat{- \\ - \\} - \end{tabular} & HOW TO HASH?? & \begin{tabular}{@{}c@{}} 0.5160 \\ 0.5158 \\ \eat{0.5081 \\ 0.9895 \\} 0.3757 \end{tabular} & \begin{tabular}{@{}c@{}} 0.5807 \\ 0.5421 \\ \eat{0.5455 \\ 0.5053 \\} 0.5415 \end{tabular} & \begin{tabular}{@{}c@{}} 0.5904 \\ 0.5632 \\ \eat{0.5556 \\ 0.6316 \\} 0.5611 \end{tabular} & \begin{tabular}{@{}c@{}} 0.6265 \\ 0.5895 \\ \eat{0.5934 \\ 0.5684 \\} 0.5893 \end{tabular} & \begin{tabular}{@{}c@{}} 0.6652 \\ 0.6632 \\ \eat{0.6396 \\ 0.7474 \\} 0.6608 \end{tabular} & \begin{tabular}{@{}c@{}} \cellcolor{gray!25}0.7703 \\ 0.7105 \\ \eat{0.6818 \\ 0.7895 \\} 0.7087 \end{tabular} & \begin{tabular}{@{}c@{}} - \\ - \\ \eat{- \\ - \\} - \end{tabular} & \begin{tabular}{@{}c@{}} - \\ - \\ \eat{- \\ - \\} - \end{tabular} & \begin{tabular}{@{}c@{}} 0.5212 \\ 0.5211 \\ \eat{0 \\ 0 \\} 0.3334 \end{tabular} & \begin{tabular}{@{}c@{}} 0.7584 \\ 0.7211 \\ \eat{0.7100 \\ 0.7474 \\} 0.7209 \end{tabular} & { \begin{tabular}{@{}c@{}} 0.7609 \\ \cellcolor{gray!25}0.7268 \\ \eat{0.7647 \\ 0.6842 \\} \cellcolor{gray!25}0.7271 \end{tabular}} & { \begin{tabular}{@{}c@{}} 0.7380 \\ 0.6737 \\ \eat{0.6667 \\ 0.6947 \\} 0.6735 \end{tabular}} \\\midrule

		\rotatebox[origin=c]{90}{digg}
		& \begin{tabular}{@{}r@{}}\textbf{AUC} \\ \textbf{ACC} \\ \eat{\textbf{Precision} \\ \textbf{Recall} \\} \textbf{F1}\end{tabular}
		& OOM & HOW TO HASH?? & \begin{tabular}{@{}c@{}} 0.5001 \\ 0.5001 \\ \eat{0.7500 \\ 0.0000 \\} 0.3338 \end{tabular} & \begin{tabular}{@{}c@{}} 0.7909 \\ 0.7751 \\ \eat{0.7515 \\ 0.8220 \\} 0.7746 \end{tabular} & \begin{tabular}{@{}c@{}} 0.7607 \\ 0.7039 \\ \eat{0.7052 \\ 0.7006 \\} 0.7039 \end{tabular} & \begin{tabular}{@{}c@{}} 0.7599 \\ 0.7030 \\ \eat{0.7004 \\ 0.7095 \\} 0.7030 \end{tabular} & \begin{tabular}{@{}c@{}} 0.7203 \\ 0.6357 \\ \eat{0.7154 \\ 0.4508 \\} 0.6228 \end{tabular} & OOT
		& OOM
		& \begin{tabular}{@{}c@{}} - \\ - \\ \eat{- \\ - \\} - \end{tabular} & \begin{tabular}{@{}c@{}} 0.5000 \\ 0.5000 \\ \eat{0 \\ 0 \\} 0.3332 \end{tabular} & \begin{tabular}{@{}c@{}} \cellcolor{gray!25}0.8185 \\ \cellcolor{gray!25}0.7982 \\ \eat{0.7453 \\ 0.9060 \\} \cellcolor{gray!25}0.7958 \end{tabular} & \begin{tabular}{@{}c@{}} 0.7611 \\ 0.7418 \\ \eat{ 0.7185 \\ 0.7952 \\} 0.7411 \end{tabular} & \begin{tabular}{@{}c@{}} 0.7587 \\ 0.7444 \\ \eat{0.7171 \\ 0.8071 \\} 0.7433 \end{tabular} \\
		
		\bottomrule
		
	\end{tabular}
	\vspace{-0.1cm}
\end{table*}

\begin{figure}[b!]
	\captionsetup[subfigure]{justification=centering}
	\centering
	\vspace{-0.5cm}
	\begin{subfigure}[t]{0.31\textwidth}
		\centering
		\includegraphics[width=\textwidth]{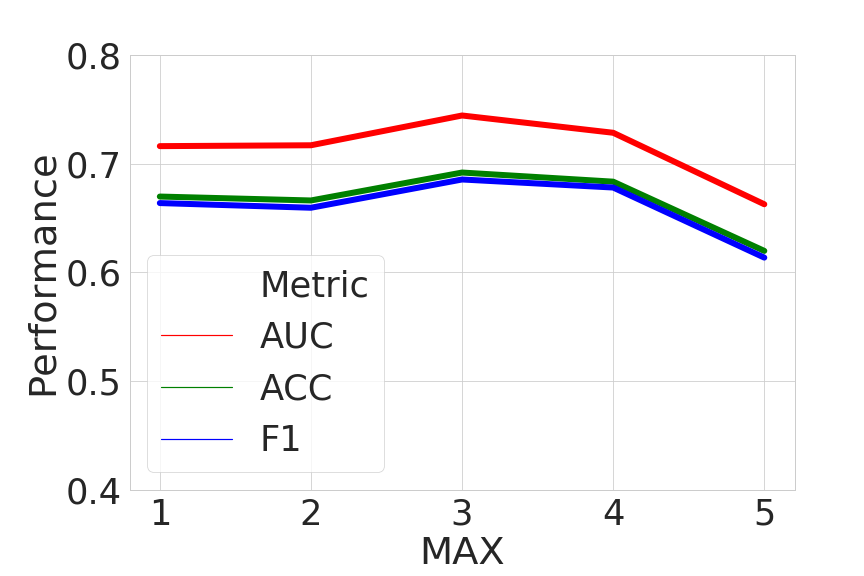}
		\caption{varying $\text{MAX}$ temporal distance $\deltat$ }
		\label{fig_sens_analysis_a}
	\end{subfigure}
	~
	\begin{subfigure}[t]{0.31\textwidth}
		\includegraphics[width=\textwidth]{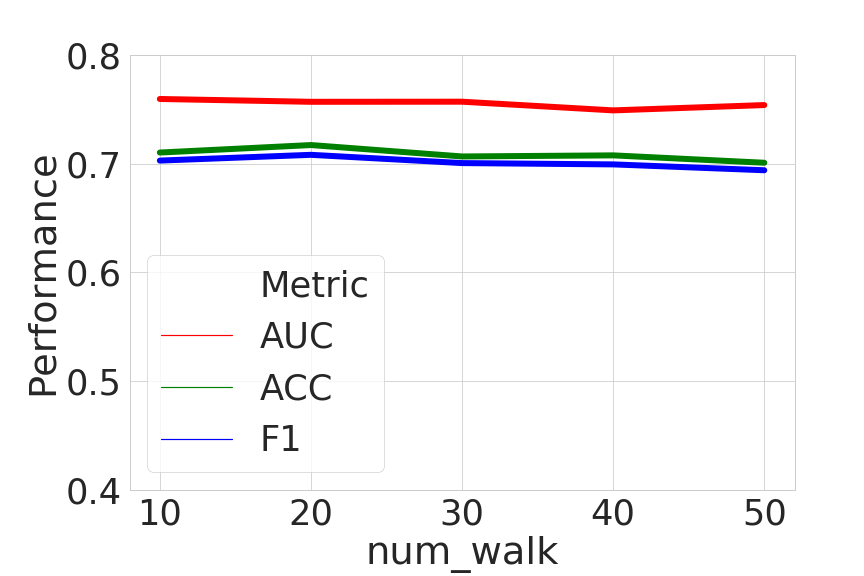}
		\caption{varying \# walks}
		\label{fig_sens_analysis_b}
	\end{subfigure}
	~
	\begin{subfigure}[t]{0.31\textwidth}
		\includegraphics[width=\textwidth]{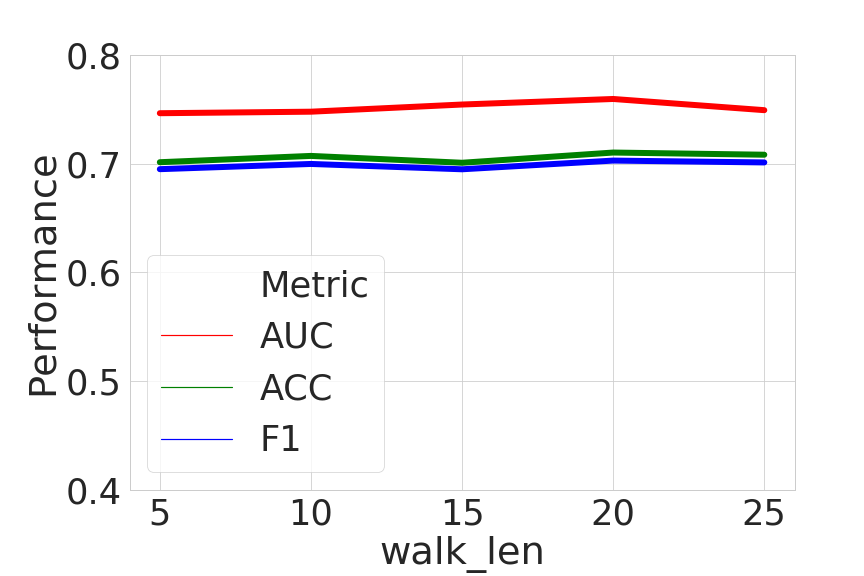}
		\caption{varying walk length}       \label{fig_sens_analysis_c}
	\end{subfigure}
	\caption{Sensitivity Analysis on bitcoin dataset. \method achieves highest scores in AUC, ACC and $F$-1 score when $\text{MAX}=3$. Increasing the numbers of walks or increasing their lengths do not significantly affect the performance of \method.}
	\label{fig_sens_analysis}
	\vspace{-.45cm}
\end{figure}

\subsection{Sensitivity analysis} \label{sec:sens_analysis}

We also perform sensitivity analysis of the hyperparameters used in this work on the bitcoin dataset. Particularly, we perform grid analysis by varying (1) max temporal distances, (2) the number of temporal walks per edge and (3) the length of walks. The results are given in Figure~\ref{fig_sens_analysis}. 
Figure~\ref{fig_sens_analysis_a} indicates that when $\text{MAX}=3$, \method achieves the best performance. This implies that although it is potentially beneficial to incorporate nodes in temporally distant contexts, it will also incorporate information that is less relevant. Therefore, we set $\text{MAX}=3$ by default for the experiments in this work.
Figure~\ref{fig_sens_analysis_b}-\ref{fig_sens_analysis_c} imply that the performance of \method is not significantly affected by the number of walks performed or the length of these temporal walks. This is reasonable because \method leverages these temporal walks to collect node features into the context and normalizes their occurrences in the histograms. Thus, adding more nodes in the ordered temporal contexts does not provide extra useful information. We empirically set the number of walks per edge to be $10$ and the lengths to be $20$ in the experiments of this work.

\subsection{Output storage and runtime in detail}
\label{sec:detailed_output_space}

We report detailed output storage in Table~\ref{table:space} and the time elapsed when running all methods in Table~\ref{table:runtime-perf}.
It can be seen that the node-wise sparse binary vectors generated by \method take trivial amount of storage compared to the other methods, while its runtime is comparable to node2vec. \method finished running on all datasets while most baselines fail to finish within the time limit on the large datasets, digg and wiki.

\begin{table}[h!]
	
	\centering
	{\scriptsize
		\vspace{-0.6cm}
		\caption{
			Space required to store the output in MB.
			\method requires $63\times$-$339\times$ less space than other embedding methods. $`-'$ indicates that the method does not apply to that dataset, or encounters errors such as out-of-memory or out-of-time.
		}
		\vspace{1mm}
		\label{table:space}
		\setlength{\tabcolsep}{2pt} \begin{tabular}{@{}r Hccccccccc|c }
			\toprule
			datasets & {\bf  Cosum } & {\bf SC} & {\bf LINE} & {\bf DW} & {\bf n2v} & {\bf  CTDNE } & {\bf  s2vec} & {\bf  DNGR } & {\bf  m2vec } & {\bf AspEm} & { \abbre} \\
			\midrule
			
			{\bf citeseer} & 1 & 6.3 & 6.4 & 6.7 & 6.6 & - & 6.9 & 7.4 & 3.2 & 6.4 & \cellcolor{gray!25}{0.033} \\
			{\bf yahoo} & - & 134.4 & 134.3 & 167.8 & 167.6 & - & - & - & 127.7 & 134.3 & \cellcolor{gray!25}{2.1} \\
			{\bf bitcoin} & - & 4.8 & 5.3 & 6.0 & 6.0 & 6.4 & 5.8 & 5.3 & - & 4.9 & \cellcolor{gray!25}{0.041} \\
			{\bf digg} & - & 369.1 & 370.0 & 469.8 & 469.8 & 486.6 & - & - & - & 369.3 & \cellcolor{gray!25}{2.9} \\
			{\bf wiki} & - & - & 1430 & - & - & 1980 & - & - & - & 1430 & \cellcolor{gray!25}{4.2} \\
			
			\bottomrule
		\end{tabular}
	}
	\vspace{-0.2cm}
\end{table}

\begin{table}[h!]
	\centering
	{\scriptsize
		\vspace{-1cm}
		\caption{Comparison between \method and baselines in terms of runtime (in seconds). Note the runtime of dynamic \method (short-term) for the \textit{temporal} networks is shown in parentheses.
		}
		\vspace{1mm}
		\label{table:runtime-perf}
		\setlength{\tabcolsep}{5pt} \def\arraystretch{0.9} \begin{tabular}{@{}cl ccccc HHHHH}
			\toprule
			&   & {\bf citeseer} & {\bf yahoo} & {\bf bitcoin} & {\bf digg} & {\bf wiki}  \\
			\midrule
			& SC    &23.72  &766.42 &4.80   &8091.09&   1   \\
			& LINE$^{*}$  &144.94 &223.87 &134.48 &   227.28   &   415.00  \\
			& DW$^{*}$    &8.90  &209.72 & 16.99 &2115.86&   -  \\
			& n2v$^{*}$   &7.99  &222.14 & 15.91 &2751.91&   -  \\
			& CTDNE    &   -   &   -   &   13.25   &2227.66&4217.19   \\
			& s2vec$^{*}$ &325.38 &   -   &897.2 &   -   &   -   \\
			& DNGR    &128.63 &   -   &97.09  &   -   &   -   \\
			& m2vec   &125.98 &   -   & -  &   -   &   -   \\
			& AspEm   &0.62  & 4.70  &0.71   &15.318 &386.24   \\
			& CN   &0.58  & 19.59  &0.70   &    63.95   & 109.11   \\
			\midrule
			& \abbre &13.15  &   221.84   &\begin{tabular}{@{}c@{}}20.52 \\ (39.97)\end{tabular}  &\begin{tabular}{@{}c@{}}1507.95 \\ (3062.13)\end{tabular} &\begin{tabular}{@{}c@{}} 1537.24\\(3997.85)\end{tabular}\\
			
			\bottomrule
		\end{tabular}
	}
	\vspace{-0.25cm}
\end{table}

\vspace{-0.2cm}
\section{Additional Related Work}
\label{app:related}

In this section we provide additional related work, complementing our discussion in \S~\ref{sec:related-work}.

\vspace{-0.2cm}
\subsubsection{Node Embeddings.} Here we give some more details about proximity-based methods, which we employ in our experiments. DeepWalk~\citelatex{deepwalk} and node2vec~\citelatex{node2vec} leverage vanilla and 2-order random walk, respectively, to explore the identities of the neighborhood; 
LINE~\citelatex{line} can be seen as a special case of DeepWalk by setting the context to be 1~\citelatex{qiu2018network}; metapath2vec~\citelatex{dong2017metapath2vec} relies on predefined meta-schema to perform random walk in heterogeneous networks.  
In the field of temporal network embedding, most approaches~\citelatex{zhu2016scalable,hisano2018semi} approximate the dynamic network as discrete static snapshots overtime, which does not apply to user stitching tasks as sessions corresponding to the same user could occur in multiple timespans.
CTDNE~\citelatex{nguyen2018continuous} first explores temporal proximity by learning temporally valid embeddings based on a corpus of temporal random walks. 
Another related field is hashing-based embedding, for example, node2hash~\cite{wang2018feature} proposes to hash the pairwise node proximity derived from random walk into low-dimensional hashcode as the embeddings. Due to the quadratic complexity in computing the pairwise proximity between nodes, node2hash does not apply to large-scale networks.
One limitation of these methods is that training a skip-gram architecture on the entire corpus sampled by random walks can be memory-intensive. 
A further limitation of these approaches, as well as existing deep architectures~\citelatex{sdne,graphsage} is that for nodes to have similar embeddings, they must be in close proximity (e.g. neighbors or nodes with several common neighbors) in the network.  This is not necessarily the case for user stitching, where corresponding entities may exhibit similar behavior (resulting in similar local topologies) but not connect to the same entities.  

Compared with proximity-based methods, embedding works exploring structural equivalency or similarity~\citelatex{role2vec,ribeiro2017struc2vec,rossi2015roles,DonnatZHL18graphwave,xnetmf,JinRKKRK19multilens,JinHSWWSK19ember,LiuZSGK19} are more suitable to handle user stitching. Representative examples include the following: 
struc2vec~\citelatex{ribeiro2017struc2vec}, xNetMf~\citelatex{xnetmf}, and EMBER~\citelatex{JinHSWWSK19ember} define similarity in terms of degree sequences in node-centric subgraphs; DeepGL~\citelatex{rossi2018deep} learns deep relational functions applied to degree, triangle counts and other graph invariants in an inductive scheme. Role2vec~\citelatex{role2vec} proposes a framework that inductively learns structural similarity by introducing attributed random walk atop relational operators, while MultiLENS~\citelatex{JinRKKRK19multilens} summarizes node embeddings obtained by recursive application of relational operators. CCTN~\citelatex{LiuZSGK19} embeds and clusters nodes in a network that are not only well-connected but also share similar behavioral patterns (\eg, similar patterns in the degree or other structural properties over time).

\vspace{-0.2cm}
\subsubsection{Locality sensitivity hashing (LSH).} More recently, LSH functions that are robust to distortion~\citelatex{aghazadeh2017rhash}; require less storage of the hash codes~\citelatex{li2011theory,weiss2009spectral}; generate codewords with balanced amounts of items~\citelatex{kang2012robust} or compute hash functions efficiently~\citelatex{li2006very,li2012one,ji2013min,shrivastava2014densifying} 
have attracted much attention. 
LSH has been used in a variety of data mining applications, including network alignment~\citelatex{heimann2018hashalign}, network inference~\citelatex{SafaviSK17}, anomaly detection~\citelatex{ManzoorMA16}, and more. 
In addition, there are also works devoted to 
learning to hash~\citelatex{aghazadeh2017rhash} where the main idea is to learn hash codes through an optimization objective function, or intelligently probe multiple adjacent code words that are likely to contain query results in a hash table for similarity search~\citelatex{lv2007multi}. But these methods do not apply to large-scale graphs directly.

\bibliographystylelatex{splncs04} \bibliographylatex{paper}

\end{document}